\documentclass{cocv}
\usepackage{graphics,float}
\usepackage[T1]{fontenc} 
\usepackage{lmodern} 
\usepackage[english]{babel} 
\usepackage{tikz}
\usetikzlibrary{arrows,shadows}
\usepackage{color}

\usepackage[pdftex]{hyperref} 

\usepackage{amsfonts,latexsym,amssymb} 
\usepackage{mathrsfs,amsmath}
\usepackage[all]{xy}
\graphicspath{{figures/}}

\usepackage{xcolor} 

\newcommand{\Proof}{{\noindent {\bf Proof:   }}}
\newcommand{\EndProof}{{\hfill $\Box \qquad $ \endtrivlist}\par \vspace{0.5cm}}%
\renewcommand{\1}{ \mbox{ \usefont{U}{bbm}{m}{n}  \selectfont 1}}

\renewcommand{\hat}{\widehat}

\newcommand{\vecE}{\mathbf{E}}

\newcommand{\vecL}{\mathbf{L}}

\newcommand{\vecO}{\mathbf{O}}
\newcommand{\vecP}{\mathbf{P}}

\newcommand{\veca}{\mathbf{a}}
\newcommand{\vecb}{\mathbf{b}}
\newcommand{\vecc}{\mathbf{c}}
\newcommand{\vecd}{\mathbf{d}}
\newcommand{\vece}{\mathbf{e}}
\newcommand{\vecf}{\mathbf{f}}
\newcommand{\vecg}{\mathbf{g}}

\newcommand{\vecp}{\mathbf{p}}
\newcommand{\vecq}{\mathbf{q}}

\newcommand{\vecs}{\mathbf{s}}

\newcommand{\vecv}{\mathbf{v}}

\newcommand{\Div}{\mbox{ div }}

    \newtheorem{theorem}{Theorem}[section]

    \newtheorem{proposition}[theorem]{Proposition}

    \newtheorem{definition}{Definition}[section]

    \newtheorem{remark}{Remark}[section]
    
    \numberwithin{equation}{section}

\begin{document}
\title{Swim-like motion of bodies immersed in an ideal fluid}
\author{Marta Zoppello}\address{Dip. di Matematica, Universit\`a di Padova, via Trieste 63, 35121 Padova, Italy, email: mzoppell@math.unipd.it }
\author{Franco Cardin}\address{Dip. di Matematica, Universit\`a di Padova, via Trieste 63, 35121 Padova, Italy, email: cardin@math.unipd.it }
%
%
\begin{abstract} 
The connection between swimming and control theory is attracting increasing attention in the recent literature. Starting from an idea of Alberto Bressan \cite{Bressan07} we study the system of a planar body whose position and shape are described by a finite number of parameters, and is immersed in a 2-dimensional ideal and incompressible fluid in terms of gauge field on the space of shapes. We focus on a class of deformations measure preserving which are diffeomeorphisms whose existence is ensured by the Riemann Mapping Theorem. We face a crucial problem: the presence of possible non vanishing initial impulse. If the body starts with zero initial impulse we recover the results present in literature (Marsden, Munnier and oths). If instead the body starts with an initial impulse different from zero, the swimmer can self-propel in almost any direction if it can undergo shape changes without any bound on their velocity. This interesting observation, together with the analysis of the controllability of this system, seems innovative.
 \end{abstract}
%
%
\maketitle
\section*{Introduction}
\label{intro}
In this work we are interested in studying the self-propulsion of a deformable body in a fluid. This kind of systems is attracting an increasing interest in recent literature. Many authors focus on two different type of fluids. Some of them consider swimming at micro scale in a  Stokes fluid \cite{AlougesDeSimone13,AlougesHeltai13,AlougesDeSimone08,AlougesGiraldi12,LoheacTucsnak13,Tucsnak11,ShapereWilczek89}, because in this regime the inertial terms can be neglected and the hydrodynamic equations are linear. Others are interested in bodies immersed in an ideal incompressible fluid \cite{Bressan07, Tucsnak10,Khapalov13,Marsdenpesci,MunnierChambrion10}
and also in this case the hydrodynamic equations turn out to be linear.\\
We deal with the last case, in  particular we study a deformable body -typically a swimmer or a fish- immersed in an ideal and irrotational fluid. This special case has an interesting geometric nature and there is an attractive mathematical framework for it. We exploit this intrinsically geometrical structure of the problem inspired by \cite{ShapereWilczek90,ShapereWilczek89} and \cite{MasonBurdick99},   in which they interpret the system in terms of gauge field on the space of shapes.  The choice of taking into account the inertia can apparently lead to a more complex system, but neglecting the viscosity the hydrodynamic equations are still linear, and this fact makes the system more manageable. The same fluid regime and existence of solutions of these hydrodynamic equations has been studied in \cite{Tucsnak10} regarding the motion of rigid bodies. \\
We start from an early idea of Alberto Bressan \cite{Bressan07} and some unpublished developments, according to which the shape changes can be described by a finite number of parameters. These kind of systems, where the controls are precisely given by further degrees of freedom of the systems, have been first studied deeply by Aldo Bressan, see e.g. \cite{BressanAldo2,BressanAldo3,BressanAldo1}. In this framework we  show that the composed system ``fluid-swimmer'' is Lagrangian geodesic. Next, coupling this fact with some techniques developed in \cite{MunnierChambrion10}, we are able to show that the kinetic energy of the system (i.e. the Lagrangian) is \textit{bundle-like}, a concept  by Bruce Reinhart \cite{Reinhart} and introduced in control theory by Franco Rampazzo in \cite{Rampazzo}.
This leads us to express the equations of motion as linear control equations, where any quadratic term is vanishing, radically simplifying our final analysis on the system. The geometric construction of the control dynamic equations follows substantially the line of thought of \cite{Cardin,marle91}.\\
At a first glance, the deformations of the swimmer are naturally given by  diffeomorphisms, that are infinite dimensional objects. By  considering a planar setting and making use of complex analysis, as suggested in \cite{MontgomeryStokes,MunnierChambrion10} the Riemann Mapping Theorem plays a crucial role in describing the shape changes of the swimmer.
It turns out that the diffeomorphisms can be parametrized by appropriate complex converging series. In the literature other authors exploit the same way of describing the shape changes by conformal maps, for example in \cite{MontgomeryStokes} in the environment of the Stokes approximate regime or in \cite{MunnierChambrion10} in the case of an ideal and irrotational fluid, in which they take into account only a finite number of terms to represent the diffeomorphisms. We follow substantially an analogous approach to merge this idea with the setting of Alberto Bressan. The choice of using a finite number of parameters means that the kind of deformations that we consider is more restricted but still enough to describe a wide range of swimmers.\\
In order to have a more manageable system that the one in \cite{MunnierChambrion10}, we establish a connection between the use of complex and real shape parameters. We show  that, if we consider small shape changes, a well precise choice of the real and imaginary part of the shape parameters leads to obtain exactly the same deformation proposed in \cite{MasonBurdick99}, which use a rather different parametrization governed by suitable small deformation. 
Therefore we get a description of our system with a finite number of parameters/coordinates, which is useful to apply the idea of controlling the shape coordinates to steer the swimmer between two different configurations. In this environment we recover the well known \textit{Scallop Theorem} \cite{AlougesDeSimoneLefebvre08} in the case in which we suppose to have only one real shape parameter.
Thanks to the idea of using a finite number of parameters we can reduce our dynamic equations to a control system. The controllability issue has been recently linked to the problem of swimming \cite{AlougesDeSimone13,AlougesDeSimoneLefebvre08,Khapalov07,LoheacTucsnak13}  since it helps in solving effectively motion planning or optimal control problems. 

We point out that in the Stokes regime there are interesting results, for example in \cite{LoheacTucsnak13} the authors study the controllability of a swimmer which performs small deformations around the sphere, or in \cite{Khapalov13} in which he considers the swimming mechanism  as a `broken-line'-like structure, formed by an ordered sequence of finitely many sets. Even in \cite{AlougesDeSimone13,GiraldiMartinon13} they study the controllability of a slender swimmer composed by $N$ links immersed in a viscous fluid at low Reynolds number. \\In the present work we deal with the ideal and irrotational case, neglecting viscosity.
In particular differently from what was done in other works, we focus our attention on a crucial problem: the presence of an initial impulse. The case of zero initial linear momentum is studied in literature concerning systems of different nature: both in the multi-particle or many-bodies field, \cite{Toshihiro98,Toshihiro99,Toshihiro01}, and shape changing bodies, \cite{Marsdenpesci,MasonBurdick99,MunnierChambrion10,ChambrionMunnier11}, as the equation of motion are a driftless affine control system whose controllability can be studied using classical techniques. Instead, the case of a non vanishing initial impulse leads us to a more complex system since the equations of motion involve also a non zero drift term and their controllability is more tricky to study. Therefore we have two contributions to the motion of the system: the first one that is purely geometrical and determined by the structure of the problem, and the second one, strictly linked to the presence of a non vanishing initial impulse.\\ 
The controllability of this kind of systems is studied in detail, and among other facts it is worth noting that we need at least three real shape parameters to make the system controllable.

We have three state parameters, three conjugate variables and at least three controls. Despite the evident  complexity of the computations linked to this number of variables, we managed to obtain interesting results.
\\\\
The plan of the paper is the following. In Section 1 we present in some detail the geometric aspects useful to formulate our problem. The proper geometrical setting of the swimmer in a 2-dimensional fluid is faced in Section 2. Section 3 contains an exhaustive study, in a complex setting, of the deformation of the body, together with the construction of the equation of motion.
We deal with all the controllability issues in Section 4.
\section{Preliminaries}

This section covers some auxiliary mathematical topics, in particular from Lie groups, fiber bundles and connections that we shall need later. This summary will be helpful to set the notation, fill in some gaps, and to provide a guide to the literature for needed background.

\subsection{Lie Groups}
Let us start from some geometric and algebraic notions on Lie groups, that arise in discussing conservation laws for mechanical and control systems and in the analysis of systems with some underlying symmetry.
\begin{definition}
A Lie group is a smooth manifold $G$ that is a group with identity element $e=g g^{-1}=g^{-1}g$, and for which the group operations of multiplication, $(g,h)\mapsto gh$ for $g,h\in G$, and inversion, $g\mapsto g^{-1}$, are smooth.
\end{definition}

Before giving a brief description of some of the theory of Lie groups we mention an important example: the group of linear isomorphisms of $\mathbb{R}^{n}$ to itself. This is a Lie group of dimension $n^{2}$ called the general linear group and denoted by $GL(n,\mathbb{R})$. The conditions for a Lie group are easily
checked. This is a manifold, since it is an open subset of the linear space of all linear maps of $\mathbb{R}^{n}$ to itself; the group operations are smooth, since they are algebraic operations on the matrix entries.

\begin{definition}
A matrix Lie group is a set of invertible $n\times n$ matrices that is closed under matrix multiplication and that is a submanifold of $\mathbb{R}^{n\times n}$.
\end{definition}

Lie groups are frequently studied in conjunction with Lie algebras, which are associated with the tangent spaces of Lie groups as we now describe.

\begin{definition}
For any pair of $n\times n$ matrices $A$, $B$ we define the matrix Lie bracket $[A,B] = AB - BA$.
\end{definition}
\begin{proposition}
The matrix Lie bracket operation has the following two properties:
\begin{itemize}
\item[(i)] For any $n\times n$ matrices $A$ and $B$, $[B,A] = -[A,B]$ (skew-symmetry).
\item[(ii)] For any $n\times n$ matrices $A$, $B$, and $C$, \\$[[A,B], C] + [[B,C],A] + [[C,A],B] = 0$. (This is known as the Jacobi identity.)
\end{itemize}
\end{proposition}
As is known, properties (i) and (ii) above are often thought as the definition of more general Lie brackets (than $AB-BA$) on vector spaces called Lie algebras.
\begin{definition}
A (matrix) Lie algebra $\mathfrak g$ is a set of $n\times n$ matrices that is a vector space with respect to the usual operations of matrix addition and multiplication by real numbers (scalars) and that is closed under the matrix Lie bracket operation $[\cdot ,\cdot]$.
\end{definition}
\begin{proposition}
For any matrix Lie group $G$, the tangent space at the identity $T_{I}G$ is a Lie algebra.\\
As usual, for matrix Lie groups one denotes $e=\mathbb{I}$
\end{proposition}

We now describe an example that plays an important role in mechanics and control.

\textbf{The plane Euclidean Group}\\
Consider the Lie group of all $3\times 3$ matrices of the form
\begin{equation}
\begin{pmatrix}
R&d\\
0&1
\end{pmatrix}
\end{equation}

where $R\in SO(2)$ and $d\in \mathbb{R}^{2}$. This group is usually denoted by $SE(2)$ and is called the special Euclidean group.
The corresponding Lie algebra, $se(2)$, is three-dimensional and is spanned by
\begin{align}
&\mathcal{A}_{1}=
\begin{pmatrix}
0&-1&0\\
1&0&0\\
0&0&0
\end{pmatrix}&&
\mathcal{A}_{2}=\begin{pmatrix}
0&0&1\\
0&0&0\\
0&0&0
\end{pmatrix}&&
\mathcal{A}_{3}=\begin{pmatrix}
0&0&0\\
0&0&1\\
0&0&0
\end{pmatrix}
\label{lie algebra}
\end{align}
The special Euclidean group is of central interest in mechanics since it describes the set of rigid motions and coordinate transformations on the plane. Let $G$ be a matrix Lie group and let $\mathfrak g = T_{I}G$ be the corresponding Lie algebra. The dimensions of the differentiable manifold $G$ and the vector space $\mathfrak g$ are of course the same, and there must be a one-to-one local correspondence between a neighborhood of $0$ in $\mathfrak g$ and a neighborhood of the identity element $I$ in $G$. An explicit local correspondence is provided by the exponential mapping $exp:\mathfrak g\mapsto G$, which we now describe. For any $A\in\mathbb{R}^{n\times n}$ (the space of $n\times n$ matrices). $exp(A)$ is defined by
\begin{equation}
exp(A):=I + A +\frac{1}{2!}A^{2} +\frac{1}{3!}A^{3} +\dots
\end{equation}
This map for $SE(2)$ can be defined by the exponential of the elements of the Lie algebra $se(2)$. More precisely
\begin{equation}
exp(\theta\mathcal{A}_1)=\begin{pmatrix}\cos\theta&-\sin\theta&0\\
\sin\theta&\cos\theta&0\\
0&0&1
\end{pmatrix}
\end{equation}
\begin{equation}
\begin{aligned}
\exp(x\mathcal{A}_2)=\begin{pmatrix}1&0&x\\0&1&0\\0&0&1\end{pmatrix}&&\exp(y\mathcal{A}_2)=\begin{pmatrix}1&0&0\\0&1&y\\0&0&1\end{pmatrix}
\end{aligned}
\end{equation}
Since $[\mathcal{A}_i,\mathcal{A}_j]=0$ for all $i,j=1,2,3$, we have that $\forall (\theta,x,y)\in\mathbb{R}^3\equiv\mathfrak{g}=se(2):$
$$
\exp(\theta\mathcal{A}_1+x\mathcal{A}_2+y\mathcal{A}_3)=\exp(\theta\mathcal{A}_1)\exp(x\mathcal{A}_2)\exp(y\mathcal{A}_3)
$$
that is clearly elements of $SE(2)$.\\
\indent We now define the action of a Lie group $G$ on a manifold $Q$. Roughly speaking, a group action is a group of transformations of $Q$ indexed by elements of the group $G$ and whose composition in $Q$ is compatible with group multiplication in $G$.
\begin{definition}
Let $Q$ be a manifold and let $G$ be a Lie group. A left action of a Lie group $G$ on $Q$ is a smooth mapping $\Phi: G\mapsto Q$ such that
\begin{itemize}
\item[(i)] $\Phi(e, q) = q $ for all $q\in Q$,
\item[(ii)]$\Phi(g, \Phi(h, q)) = \Phi(gh, q)$ for all $g, h\in G$ and $q\in Q$,
\item[(iii)]$\Phi(g,\cdot)$ is a diffeomorphism for each $g\in G$.
\end{itemize}
\end{definition}
A Lie group acts on its tangent bundle by the tangent map. We can consider the left or the right action of $G$ on $\mathfrak{g}$ by: $T_{e}L_{g}\xi$ or $T_{e}R_{g}\xi$, where $L_{g}$ and $R_{g}$ denote left and right translations, respectively; so if $g=g(t)$ is a curve in $G$, then there exists a time dependent $\xi(\cdot)\in\mathfrak g$ such that
\begin{equation}
\label{group_diff}
\dot{g}(t)=T_eL_{g(t)}\xi(t)=g(t)\xi(t)
\end{equation}
and similarly for the right action.\\
Given left action of a Lie group $G$ on $Q$, $\Phi:G\times Q\rightarrow Q$, and $\xi$ an element of the Lie algebra $\mathfrak g$ then $\Phi^{\xi}:\mathbb{R}\times Q\rightarrow Q:(t,q)\longmapsto \Phi(\exp t\xi, q)$ is a flow on $Q$, the corresponding vector field on $Q$ is called \textbf{infinitesimal generator} of $\Phi$ corresponding to $\xi$, is denoted by $\xi_{Q}(q)$
\begin{equation}
\label{infinitesimal_generator}
\xi_{Q}(q)=\frac{d}{dt}\Phi(\exp t\xi,q)|_{t=0}\,.
\end{equation}

\subsection{Fiber Bundles and Connections}
Fiber bundles provide a basic geometric structure for the understanding of many mechanical and control problems.

A fiber bundle essentially consists of a given space (the base) together with another space (the fiber) attached at each point, plus some compatibility conditions. More formally, we have the following:
\begin{definition}
Let $\mathcal{S}$ be a differentiable base manifold and $G$ a Lie group. A differentiable manifold $Q$ is called \textsl{principal fiber bundle} if the following conditions are satisfied:
\begin{itemize}
\item[1]
$G$ acts on $Q$ to the left, freely and differentiably:
\begin{equation}
\Phi:G\times Q\rightarrow Q\\
\end{equation}
writing $\Phi(g,q)=\Phi_g\cdot q= g\cdot q\,.$
\item[2]
$\mathcal{S}=Q/G$ and the canonical projection $\pi:Q\rightarrow\mathcal{S}$ is differentiable
\item[3]
$Q$ is locally trivial, namely every point $s\in\mathcal{S}$ has a neighborhood $U$ such that $\pi^{-1}(U)\subset Q$ is isomorphic to $U\times G$, in the sense that $q\in \pi^{-1}(U)\mapsto(\pi(q),\phi(q))\in U\times G$ is a diffeomorphism such that $\phi:\pi^{-1}(U)\rightarrow G$ satisfies $\phi(g\cdot q)=g\phi(q),\forall g\in G$
\end{itemize}
\label{pr_fiber_bundle}
\end{definition}

An important additional structure on a bundle is a \textbf{connection}. Suppose we have a bundle and consider (locally) a section of this bundle, i.e., a choice of a point in the fiber over each point in the base. We call such a choice a ``field''. The idea is to single out fields that are ``constant''. For vector fields on a linear space, for example, it is clear what we want such fields to be; for vector fields on a manifold or an arbitrary bundle, we have to specify this notion. Such fields are called ``horizontal''. A connection is used to single out horizontal fields, more precisely fields which live in a subspace of the the tangent space, and is chosen to have other desirable properties, such as linearity.

\begin{definition}
Let $(Q,\mathcal{S},\pi,G)$ be a principal fiber bundle. the kernel of $T_{q}\pi$ denoted by $V_{q}:=\left\{v\in T_{q}Q|T_{q}\pi(v)=0\right\}$, is the subspace of $T_{q}Q$ tangent to the fiber through $q$ and is called \textsl{vertical subspace}. A \textsl{connection} on the principal fiber bundle is a choice of a tangent subspace $H_{q}\subset T_{q}Q$ at each point $q\in Q$ called \textsl{horizontal subspace} such that:
\begin{itemize}
\item[(1)]
$T_{q}Q=H_{q}\oplus V_{q}$
\item[(2)]
For all $g\in G$ and $q\in Q$, $T_{q}\Phi_{g}\cdot H_{q}=H_{g\cdot q}$
\item[(3)]
$H_{q}$ depends differentiably on $q$
\end{itemize}
\label{connessione}
\end{definition}
Hence, for any $q \in Q$, we have that $T_q \pi$ determines an isomorphism $H_q \cong T_{\pi(q)} \mathcal{S}$: for all $T_q Q\ni v=v_{V_{q}}+v_{H_{q}}$ and we have that $T_{\pi(q)}(v)=v_{H_{q}}\in\mathcal{S}$. In other words the choice of an horizontal subspace can be seen also as the choice of a vector valued ``connection one form'' which vanishes on the horizontal vectors.

It follows the definition
\begin{definition}
\label{Ehresmann_connection}
An Ehresmann connection A is a vector valued one form such that
\begin{itemize}
\item [(i)]
A is vertical valued: $A_q:T_q\longrightarrow V_q$ is a linear map for each point $q\in Q$
\item [(ii)]
A is a projection: $A(v)=v$ for all $v \in V_q$.
\end{itemize}
\end{definition}
In the special case in which $(Q,\mathcal{S},\pi,G)$ is a principal fiber bundle the previous conditions on $A:TQ\longrightarrow \mathfrak g$ read:
\begin{itemize}
\item[(i)]
$A(\xi_Q(q))=\xi$ for all $\xi\in\mathfrak g$ and $q\in Q$, where $\xi_Q(q)$ is the infinitesimal generator of the left action of $G$ on $Q$ (\ref{infinitesimal_generator}).
\item[(ii)]
$A$ is equivariant:
$$
A(T_q(\Phi_g(v)))=Ad_g(A(v))
$$
for all $v\in T_q Q$ and $g\in G$ where $\Phi_g$ denotes the given action of $G$ on $Q$ and where $Ad$ denotes the adjoint action of $G$ on $\mathfrak g$ defined as
$$
Ad_g:=T_e(L_g\circ R_{g^{-1}}):\mathfrak{g}\rightarrow \mathfrak{g}\,.
$$
\end{itemize}

Therefore it is evident that the horizontal subspace $H_q$ is the kernel of $A_q$.\\
n the case in which there is a metric $h(q)$ in our manifold $Q$, we have a special way to define the horizontal subspace: it is the orthogonal with respect to the metric to the vertical subspace.
\begin{equation}
H_q=\{w\in T_qQ: \left\langle w,h (q)v\right\rangle=0,\forall v\in V_q\}
\end{equation}
 In this special case our connection $A$ is called \textit{mechanical connection} (see \cite{MarsdenLect} and therein references).
We now would like to express the connection in coordinates, in order to do this we first introduce the following definition
\begin{definition}
Let us consider the following diagram\
\begin{figure}[H]
\begin{center}
\begin{tikzpicture}[scale=1]
 \draw (0,0) node[above ] {$Q$};
 \draw[->] (0,0)--(0,-2.5);
 \draw (0,-2.5) node[below] {$\mathcal{S}$};
\draw (0.5,-2.8) node{$\supset$};
\draw (1,-2.8) node{$U$};
\draw[<-] (0.1,0)--(1,-2.6);
\draw (0,-1.7) node[above left] {$\pi$};
\draw (0.7,-1.7) node[above right] {$\sigma$};
\draw (1.4,-1.7) node[above right] {where $\pi\circ\sigma=id|_{U}$};
\end{tikzpicture}
\end{center}
\end{figure}
The functions like $\sigma$ are \textbf{sections} and we call $\Gamma(U,Q)$ the set of all sections from $U$ in $Q$.
\end{definition}
Alternatively often a connection is introduced as a derivation $\nabla$ as follows.
Let $\nabla$ be a map
\begin{equation*}
 \begin{aligned}
&\nabla:\Gamma(Q)\rightarrow\Gamma(Q\otimes T^*\mathcal{S})\quad\text{such that}\\
&\nabla(\sigma_1+\sigma_2)=\nabla(\sigma_1)+\nabla(\sigma_2)\\
&\nabla(f\sigma)=f\nabla(\sigma)+\sigma\otimes df\quad \text{if $f$ is a $C^{\infty}$ function.}
\end{aligned}
\end{equation*}
Let now $\vece$ be a local basis of sections of the principal fiber bundle, in this basis the connection one-form $A$ can be expressed as
$$
e_a A^a_{b}=\nabla e_{b}\quad\,\,a,b=1\cdots dim(Q).
$$
If we change basis in $\Gamma(Q)$, say $\vece=\tilde{\vece}\Omega$, the connection $A$ changes, i.e.
\begin{align*}
\tilde{\vece}\tilde{A}=&\nabla\tilde{\vece}=\nabla(\vece\Omega^{-1})=(\nabla\vece)\Omega^{-1}+\vece d\Omega^{-1}=\vece A\Omega^{-1}+\vece d\Omega^{-1}\\
&=\tilde{\vece}\Omega A\Omega^{-1}+\tilde{\vece}\Omega d\Omega^{-1}
\end{align*}

therefore $A$ and $\tilde A$ satisfy the following relation
\begin{equation}
\label{transformation_law}
\tilde A=\Omega A\Omega^{-1}+\Omega d\Omega^{-1}
\end{equation}

Let $u(t)$ be a smooth curve in $\mathcal{S}$ passing through the point
$P = u(0)$. Let $q \in Q_P = \pi^{-1}(P)$ be any point in the fiber
of $Q$ over $P$. We would like to find a smooth curve $ \gamma(t)$
in $Q$ such that $\pi( \gamma(t)) = u(t)$, $\gamma(0) = q$,
and $ \gamma'(t) \in H_{ \gamma(t)}$ (i.e., the tangent vectors
to the curve $ \gamma(t)$ are horizontal).

From the usual theory of differential equations it follows that such a curve
$\gamma(t)$ exists and is unique, at least locally at any point $q \in Q$
(i.e., for small values of $t$). The curve $\gamma$ is called a
\textbf{horizontal lift} of $u$.
Regarding the tangent vectors, for any $q \in Q$ and any vector $\dot{u} \in T_{\pi(q)} \mathcal{S}$ there exists a unique
vector $v \in H_q \subset T_q Q$ such that $T_q \pi: v \mapsto \dot{u}$.
The vector $v$ is called the \textbf{horizontal lift} of $\dot{u}$.

Given an Ehresmann connection we can define the horizontal lift of curves in $\mathcal{S}$,
hence we can also define a notion of parallel transport that allows us
to identify different fibers of $Q$.
\begin{figure}[H]
	\begin{center}
		\includegraphics[width=9cm]{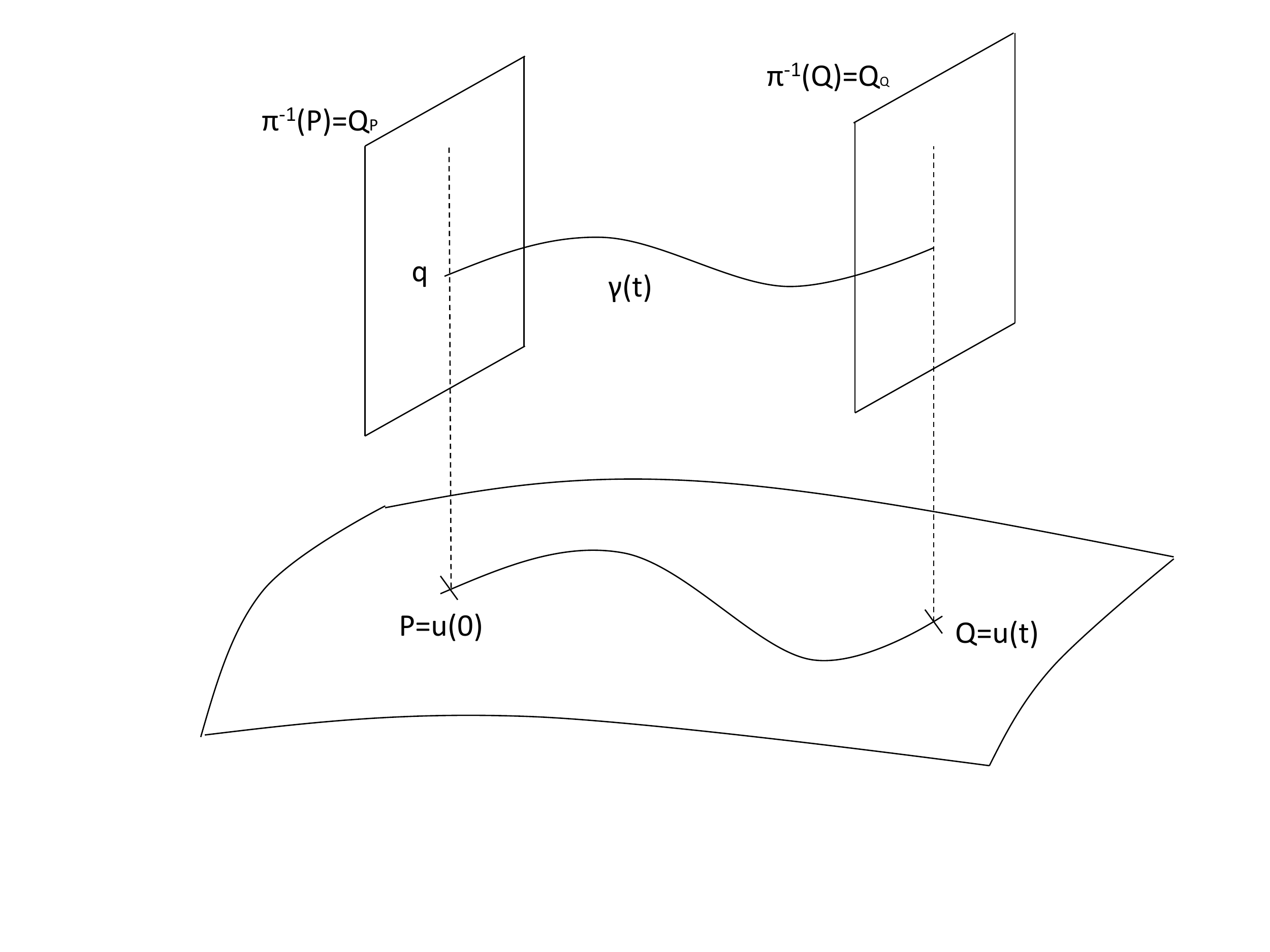}
		\caption{Horizontal lift of the curve $u(t)$}
	\label{fig:Horizontal_lift}
	\end{center}
\end{figure}

Note that, in general, the parallel transport will be path-dependent.
If we start with two different curves $u_1(t)$ and $u_2(t)$,
such that $u_1(0) = u_2(0) = P$ and $u_1(\bar{t}) = u_2(\bar{t}) = S$,
then the horizontal lifts $\gamma_1$ and $\gamma_2$ at a point $q \in Q_P$
will not meet, in general, at a point in the fiber $Q_S$, i.e., we will have
$\gamma_1(\bar t) \neq \gamma_2(\bar t)$. This gap on the fiber is called \textbf{holonomy} and depends on the choice of the connection and on the topology of the base manifold. In particular if it is connected the holonomy depends on the basepoint  only up to conjugation \cite{Kobayashi}.
\section{Geometrical setting}
In this section we present the geometrical framework underlying dynamical control systems. We derive the equations of motion and discuss how to use the geometrical tools introduced before to gain informations on our system.
\subsection{Geometry of control equations}
In this subsection we derive the local dynamic equations for the control system $(Q,h,\mathcal{F})$ where $\mathcal{F}$ is a smooth $k$-dimensional foliation on $Q$, and $h$ is the Riemaniann metric on the manifold $Q$, as done in \cite{marle91}. As is well known, on a set $U\subset Q$ adapted for the foliation, $\mathcal{F}$ coincides with the model foliation of $\mathbb{R}^n$ by $k$-dimensional hyperplanes. 
Let $\phi:U\longrightarrow \mathbb{R}^n$, $\phi(P)=(x,y)$ be a local chart of $Q$ in $U$, distinguished for $\mathcal{F}$, so that $\phi$ maps $\mathcal{F}|_U$ into the trivial fibration $\pi(x,y)=y$. Set $q=(x,y)\in Q$; given a path $u(t)\in \pi(\phi(U))$, we suppose that for every $t$, the reaction forces that implement the (ideal) constraint $y\equiv u(t)$ are workless with respect to the set $V_{q(t)}U=\ker T_{q(t)}\pi$ of the virtual displacements compatible with the constraint $y\equiv u(t)$.  

Let $(Q,h,\mathcal{F})$ be a foliated Riemanian manifold, let $U\subset Q$ be an open set adapted for $\mathcal{F}$ an let $q=(x,y)$.  If $T(q,\dot{q})=\frac{1}{2}\dot{q}^t h(q)\dot{q}$ is the kinetic energy of the unconstrained system $(Q,h,\mathcal{F})$, then the kinetic energy of the system subject to the time dependent constraint $y\equiv u(t)$ is $T(x,u(t),\dot{x},\dot{u}(t))$. The related dynamic equations are, in Lagrangian formalism
\begin{equation}
\frac{d}{dt}\frac{\partial T}{\partial \dot{x}}-\frac{\partial T}{\partial x}=0
\end{equation}
These can be put in Hamiltonian form by performing a partial legendre transformation on the $\dot{x}$- variables.
When we identify $y$ with $u(t)$ and $\dot{y}$ with $\frac{d}{dt}u(t)$, the above Lagrange equations are equivalent to
\begin{align}
\dot{x}=\frac{\partial H}{\partial p}(x,p,u,\dot{u})&&\dot{p}=-\frac{\partial H}{\partial x}(x,p,u,\dot{u})\,.
\end{align}
We call these equations \textit{control equations}. Let
\begin{equation}
\label{kinetic_energy_gen}
\dot{q}^t h(q)\dot{q}=\dot{x}^t\mathcal{C}\dot{x}+\dot{x}^t\mathcal{M}\dot{y}+\dot{y}^t\mathcal{M}^t\dot{x}+\dot{y}^t\mathcal{B}\dot{y}
\end{equation}
be the local block representation of the metric $h$ in $\phi(U)$, where $\mathcal{C},\mathcal{B}$ are symmetric and invertible respectively $k \times k$ and $(n-k)\times(n-k)$ matrices.

To every $q\in U$ denote with $H_{q} U$ the subspace orthogonal to $V_{q} U=\ker T_{q}\pi$ with respect to $h$. Referring to the local expression of $h$ in $U$, it is easy to see that $H_q U$ is the space orthogonal to the vectors $(e_i,0)_{i=1\cdots n}$ with respect to the metric $h$ .
$$H_{q} U=\{(\dot{x},\dot{y})\in T_q U\ \text{such that}\ \mathcal{C}(q)\dot{x}+\mathcal{M}(q)\dot{y}=0\}\,.
$$
Therefore $H_q U$ can be equivalently assigned through the $V_q U$-valued connection one form defined in \ref{Ehresmann_connection}
\begin{equation}
\label{conn}
A(q)=(dx+C(q)dy)\otimes\frac{\partial}{\partial x}\qquad\text{where (see \ref{Ham1_gen} )  $C=\mathcal{C}^{-1}\mathcal{M}$}
\end{equation}
whose kernel and range are respectively $H_q U$ and $V_q U$. Now we consider the orthogonal splitting of a vector into its horizontal ad vertical components
$$
v=v^v+v^h=A(q)v+hor(T_q \pi v)=(\dot{x}+C\dot{y},0)+(-C\dot{y},\dot{y})
$$
Using the above decomposition, we get the induced splitting of the kinetic energy metric tensor into its vertical and horizontal part:
\begin{equation}
h(q)dq\otimes dq=\mathcal{C}(q)A(q)\otimes A(q)+K(q)dy\otimes dy
\end{equation}
where $K(q)=\mathcal{B}-\mathcal{M}^{t}\mathcal{C}^{-1}\mathcal{M}$.
\begin{definition}
\label{bundle_like}
The Riemannian metric $h$ is \textbf{bundle-like} for the foliation $\mathcal{F}$ iff on a neighborhood $U$ with adapted coordinates $(x,y)$ the above orthogonal splitting of $g$ holds with $K=K(y)$.
\end{definition}
The importance of this notion will be clear in the following subsection (\ref{impulse}).
Using this notation we want to rewrite the control equations.

From
\begin{equation*}
p=\frac{\partial T}{\partial \dot{x}}=\mathcal{C}\dot{x}+\mathcal{M}\dot{y}
\end{equation*}
we obtain
\begin{equation}
\label{Ham1_gen}
\dot{x}=\mathcal{C}^{-1}p-\mathcal{C}^{-1}\mathcal{M}\dot{u}=\mathcal{C}^{-1}p-C\dot{u}
\end{equation}
Substituting (\ref{Ham1_gen}) in (\ref{kinetic_energy_gen}) and recalling that $-\frac{\partial H}{\partial x}=\frac{\partial T}{\partial x}$ we have
\begin{equation}
\label{Ham2_gen}
\dot{p}=-\frac{\partial H}{\partial x}=\frac{\partial T}{\partial x}=-\frac{1}{2}p^t\frac{\partial\mathcal{C}^-1}{\partial x}p+p^t\frac{\partial C}{\partial x}\dot{u}+\frac{1}{2}\dot{u}^t\frac{\partial(\mathcal{B}-\mathcal{M}^t\mathcal{C}^{-1}\mathcal{M})}{\partial x}\dot{u}
\end{equation}
Therefore the control equations are
\begin{equation}
\label{Ham_tot}
\begin{cases}
\dot{x}=\mathcal{C}^{-1}p-\mathcal{C}^{-1}\mathcal{M}\dot{u}\\
\dot{p}=-\frac{1}{2}p^t\frac{\partial\mathcal{C}^{-1}}{\partial x}p+p^t\frac{\partial C}{\partial x}\dot{u}+\frac{1}{2}\dot{u}^t\frac{\partial(\mathcal{B}-\mathcal{M}^t\mathcal{C}^{-1}\mathcal{M})}{\partial x}\dot{u}
\end{cases}
\end{equation}
We now introduce, following \cite{marle91}, the global version of the above dynamic equations when $Q$ is the total space of a surjective submersion $\pi:Q\longrightarrow \mathcal{S}$. Let $VQ$ be the vertical subbundle and $V^{*}Q$ the dual of $VQ$. Denote with $p_{Q}:T^{*}Q\longrightarrow Q$ the cotangent projection and set $\tilde{\pi}:=\pi\circ p_Q$, $\tilde{\pi}:V^{*}Q\longrightarrow \mathcal{S}$. If $(x,y)$ are local fibered coordinates on $Q$, $(x,y,p)$ are local fibered coordinates on $V^{*}Q$. Moreover, denote with $z=(x,p)$  the local coordinates on the $\tilde{\pi}$-fiber over $y$. Now, to every $y\in \mathcal{S}$, $\tilde{\pi}^{-1}(y)$ is a fiber canonically simplettomophic to $T^{*}(\pi^{-1}(y))$, representing the phase space of the constrained system restricted to the $\pi$-fiber over $y$. \\
\begin{figure}[H]
\begin{center}
\begin{tikzpicture}[scale=1]
\draw (-2,0) node[above ] {$T^*Q$};
\draw [->](-2,0) --(-2,-1.5);
\draw (-2,-1.5) node[below] {$V^*Q$};
\draw (-2,-1.8) node[below] {$(x,y,p)$};
\draw [->](-2,-2.4) --(-0.6,-3.5);
\draw [->](-2.1,-2.4) --(-0.6,-5.5);
\draw (-0.5,-3.5) node[below] {$Q$};
\draw (-0.5,-5) node[above right] {$\pi$};
\draw (-0.5,-3.8) node[below] {$(x,y)$};
\draw (1,-1.5) node[below] {$TQ$};
\draw (-0.8,-2.5) node[below] {$p_Q$};
\draw (1,-1.8) node[below] {$(x,y,\dot{x},\dot{y})$};
\draw [->](1,-2.4) --(-0.5,-3.5);
\draw [->](-0.5,-4.4)--(-0.5,-5.5);
\draw (-0.5,-5.5) node[below] {$\mathcal{S}$};
\draw (-0.5,-6) node[below] {$y$};
\draw (-1.5,-3.5) node[below left ] {$\tilde{\pi}$};
\end{tikzpicture}
\end{center}
\end{figure}


Suppose that a control vector field  $\dot{u}$ is given on $\mathcal{S}$ and that the path $u(t)$ is an integral curve of $\dot{u}$. Then the dynamic equations (\ref{Ham1_gen}) and (\ref{Ham2_gen}) are the local expression of a vector field $D_{\dot{u}}$ over $V^{*}Q$ that projects on $\dot{u}$ by $\tilde{\pi}$. Moreover the field $D_{\dot{u}}$ is tangent to the fiber of $\tilde{\pi}$ only if the control  is trivial: $\dot{u}$ vanishing.
Let us suppose that the control is given by a curve $u:[t_1,t_2]\rightarrow \mathcal{S}$ in $\mathcal{S}$ that is the integral curve of the vector field $\dot{u}$. Thus the movement of the system is described by a differentiable curve $\gamma:[t_1,t_2]\rightarrow Q$ such that $\pi(\gamma(t))=u(t)$. Note that $\frac{d\gamma}{d t}:[t_1,t_2]\rightarrow TQ$ is the natural increase of the curve $\gamma$ in the fiber tangent to $Q$. Composing $\frac{d\gamma}{d t}$ with the Legendre transform $\mathcal{L}_TQ\rightarrow T^{*}Q$ and with the projection $\tau:T^{*}Q\rightarrow V^{*} Q$ we obtain the parametric curve $\hat{\gamma}=\tau\circ\mathcal{L}\circ\frac{d\gamma}{dt}:[t_1,t_2]\rightarrow V^{*}Q$ which represent the evolution of the system taking into account the control.

Let $hor_{Q}:T\mathcal{S}\longrightarrow TQ$ denote the horizontal lift of the Ehresmann connection, introduced in the previous section, and $p_{Q}$ the cotangent projection, using the above definitions we introduce the function
\begin{equation*}
K_{\dot{u}}:V^*Q\longrightarrow\mathbb{R}\quad \quad K_{\dot{u}}\circ p_{Q}^{-1}(q)=(hor_{Q}(q)(\dot{u}))^t h(q)hor_{Q}(q)(\dot{u})
\end{equation*}

\begin{theorem}
\label{splitting_control_eq}
To every control vector field $\dot{u}$ on $\mathcal{S}$, the corresponding dynamic vector field $D_{\dot{u}}$ can be expressed as the sum of three terms:
\begin{equation}
D_{\dot{u}}=X_{H_{0}}-X_{K_{\dot{u}}}+hor(\dot{u})
\end{equation}
with
\begin{align}
&X_{H_0}=\mathcal{C}^{-1}p\frac{\partial}{\partial x}-\frac{1}{2}p^t \frac{\partial\mathcal{C}^{-1}}{\partial x}p\frac{\partial}{\partial p}\\
&-X_{K_{\dot{u}}}=\frac{1}{2}\dot{u}^t\frac{\partial K}{\partial x}\dot{u}\frac{\partial}{\partial p}\\
&hor(\dot{u})=\bigl(\frac{\partial}{\partial y}-C\frac{\partial}{\partial x}+p^t\frac{\partial C}{\partial x}\frac{\partial}{\partial p}\bigr)\dot{u}
\end{align}
where $X_{H_{0}}$ is the Hamiltonian vector field corresponding to the case of locked control, $X_{K_{\dot{u}}}$ is the Hamiltonian vector field on $V^{*}Q$ associated to $K_{\dot{u}}$ and hor is the horizontal lift of an Ehresmann connection on $\tilde{\pi}:V^{*}Q\longrightarrow \mathcal{S}$ entirely determined by $\pi$ and the metric. These equations are exactly the control equations \eqref{Ham_tot}.
\end{theorem}
\Proof
\cite{Cardin}
\EndProof

\subsubsection{The importance of initial impulse}
\label{impulse}
In what follows let us suppose that the metric $h$ is \textbf{bundle like}. 
\begin{proposition}
The control system \eqref{Ham_tot} is of two different types depending on the value of the initial value of the $x$ conjugate variables $p$.
\begin{enumerate}
\item Case $\boxed{p(0)=0}$\\
The system  \eqref{Ham_tot}  is an affine non linear driftless control system;
\item  Case $\boxed{p(0)\neq 0}$\\
The system  \eqref{Ham_tot}  is an affine non linear control system with drift.
\end{enumerate}
\end{proposition}

\Proof
Since we have supposed to have a bundle like metric we have that

$$\frac{\partial K(y)}{\partial x}=\frac{\partial(\mathcal{B}-\mathcal{M}^t\mathcal{C}^{-1}\mathcal{M})(y)}{\partial x}=0\,.$$ 

Therefore the  control equation \eqref{Ham_tot} becomes
\begin{equation}
\label{control_system}
\begin{cases}
\dot{x}=\mathcal{C}^{-1}p-\mathcal{C}^{-1}\mathcal{M}\dot{u}\\
\dot{p}=-\frac{1}{2}p^t\frac{\partial\mathcal{C}^-1}{\partial x}p+p^t\frac{\partial C}{\partial x}\dot{u}
\end{cases}
\end{equation}
\textbf{Case $p(0)=0$}.\\
The function $p(t)=0$ is the unique solution of \eqref{control_system}$_2$ according to the Cauchy theorem. Thus \eqref{control_system}$_1$ becomes a driftless control system.
$$
\dot{x}=-\mathcal{C}^{-1}\mathcal{M}\dot{u}
$$
It is clear that this last equation is entirely determined by the connection (see \eqref{conn}). Therefore in the case of null initial impulse case only the geometry of the system determines its motion.\\
\textbf{Case $p(0)\neq 0$}.\\
In this case  the equation  \eqref{control_system}$_2$ has no trivial solution that is $p(t)\neq 0$. Thus \eqref{control_system}$_1$ is a non linear control system with drift determined exactly by the presence of a non zero $p$
$$
\dot{x}=\mathcal{C}^{-1}p-\mathcal{C}^{-1}\mathcal{M}\dot{u}
$$
The presence of the drift is crucial because in this case the motion of the system is determined both by the connection (given by the geometry) and by the impulse, that is non zero.
This proves the importance of the initial value of $p$.
\EndProof
In this work we analyze both the cases. The one with zero initial impulse is well studied in literature for many systems \cite{MunnierChambrion10,MasonBurdick99,Toshihiro98,Toshihiro99,Toshihiro01}. The one with $p(0)\neq 0$ is becoming of increasing interest since the presence of the impulse influences deeply the motion, as we have seen. We deal with this problem that is more complex and tricky to study because of the presence of the drift.
\subsection{Geometric and dynamic phase}
\label{geometric_dynamic}
As we have seen, in the general theory, connections are associated with bundle mappings, which project larger spaces onto smaller ones. The larger space is the bundle, and the smaller space is the base. Directions in the larger space that project to a point are vertical directions. The connection is a specification of a set of directions, the horizontal directions, at each point, which complements the space of vertical directions. In general, we can expect that for a horizontal motion in the bundle corresponding to a cyclic motion in the base, the vertical motion will undergo a shift, called a phase shift, between the beginning and the end of its path. The magnitude of the shift will depend on the curvature of the connection and the area that is enclosed by the path in the base space: it is exactly the holonomy. This shift in the vertical element is often given by an element of a group, such as a rotation or translation group, and is called also the \textbf{geometric phase}. Referring to what said in the previous subsection, the motion is determined only by the geometrical properties of the system if it starts with zero  initial impulse.
In many examples, the base space is the control space in the sense that the path in the base space can be chosen by suitable control inputs to the system, i.e. changes in internal shape. In the locomotion setting, the base space describes the internal shape of the object, and cyclic paths in the shape space correspond to the movements that lead to translational and rotational motion of the body.\\
Nevertheless the shape changes are not the only ones to determine a net motion of the body. More generally, this motion can always be decomposed into two components:
the geometric phase, determined by the shape of the path and the area enclosed by it, and the \textbf{dynamic phase}, driven by the internal kinetic energy of the system characterized by the impulse.
It is important to stress the difference between the two phases. The geometric phase is due entirely to the geometric structure of the system. Instead the dynamic phase is present if and only if the system has non zero initial impulse or if the impulse  is not a conserved quantity, in our context we refer to what is explained in subsection \ref{impulse}. More precisely if the curvature of the connection is null, not necessarily the system does not move after a cyclic motion in the base: a net motion can result if the system starts with non zero initial impulse, and this motion is entirely due to the dynamic phase.

\begin{figure}[H]
	\centering
		\includegraphics[width=9cm]{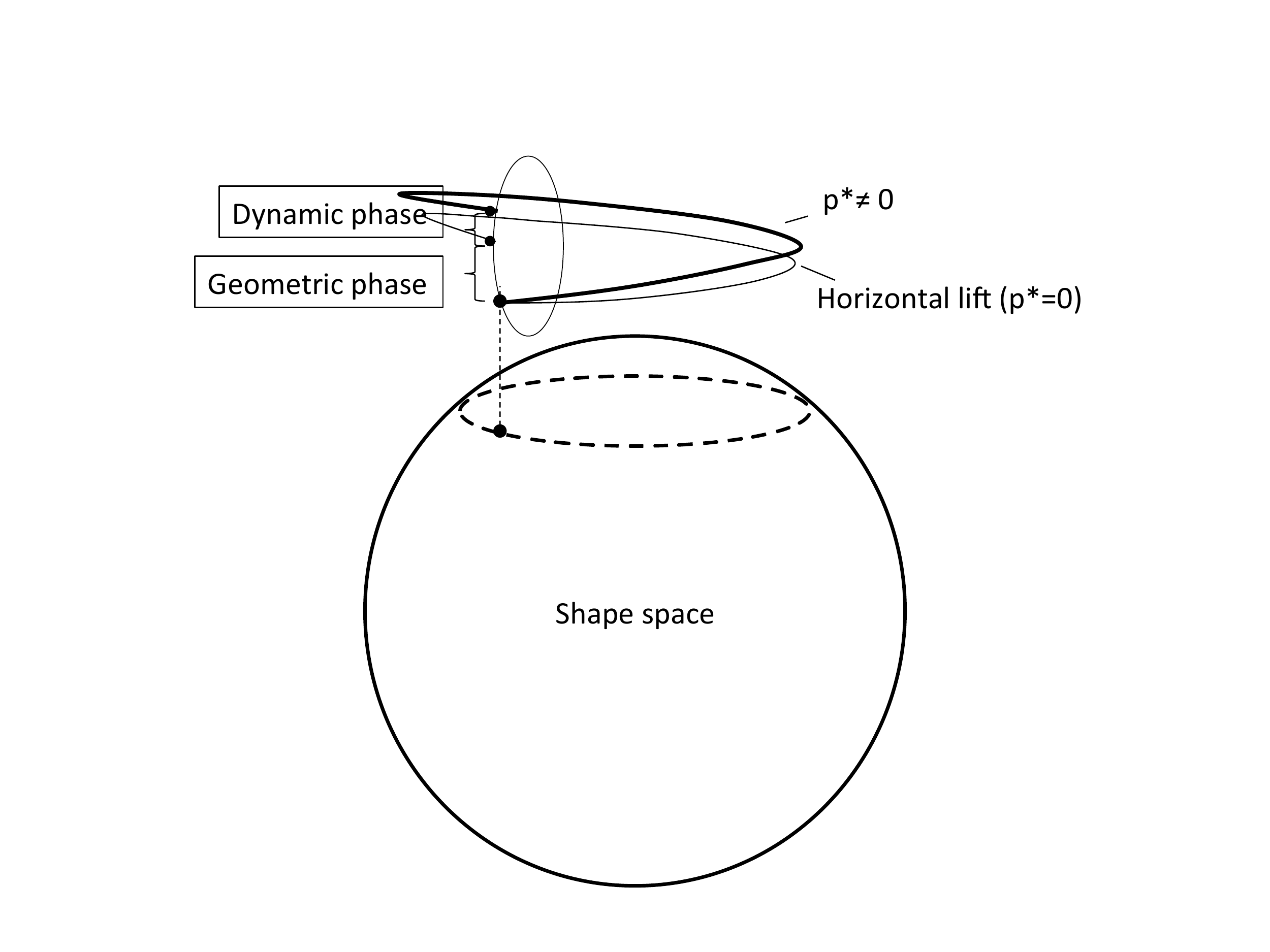}
	\caption{Geometric phase and dynamic phase.}
	\label{fig:Phase}
\end{figure}

Figure (\ref{fig:Phase}) shows a schematic representation of this decomposition for general rigid body motion. In this figure the sphere represents the base space, with a loop in the shape space shown as a circular path on the sphere. The closed circle above the sphere represents the fiber of this bundle attached to the indicated point. Given any path in the base (shape) space, there is an associated path, called the horizontal lift, that is independent of the time parametrization of the path and of the initial vertical position of the system. Following the lifted path along a loop in the shape space leads to a net change in vertical position along the fiber. This net change is just the geometric phase. On top of that, but decoupled from it, there is the motion of the system driven by the impulse, (if it is not zero) which leads to the dynamic phase. Combining these two provides the actual trajectory of the system.
\subsection{Gauge potential}
\label{Gauge_potential}
Let us consider a planar body immersed in a 2 dimensional fluid, which moves changing its shape. For the moment we do not specify the kind of fluid in which it is immersed that can be either ideal and incompressible or a viscous one with low Reynolds number. Our aim is to show that the motion of this deformable body through the fluid is completely determined by the geometry of the sequence of shapes that the idealized swimmer assumes, and to determine it. This idea was introduced by Shapere and Wilczek in  \cite{ShapereWilczek89} 
\cite{ShapereWilczek90} and developed in \cite{MontgomeryStokes}, where they apply geometrical tools to describe the motion of a deformable body in a fluid, focusing their attention on the Stokes regime.\\
The configuration space of a deformable body is the space of all possible shapes. We should distinguish between the space of shapes located somewhere in the plane and the more abstract space of unlocated shapes. The latter space can be obtained from the space \textit{cum} locations by making the quotient with the group of rigid motions in the plane, i.e declaring two shapes with different centers of mass and orientation to be equivalent. The first problem we wish to solve can be stated as follows: what is the net rotation and translation which results when a deformable body undergoes a given sequence of unoriented shapes? The problem is intuitively well posed: when a body changes its shape in some way a net rotation and translation is induced. For example, if the system is composed simply by the body, its net rigid motion can be computed by making use of the law of conservation of momentum, if instead the body is immersed in an ideal incompressible fluid this motion can be found by solving the Euler equations for the fluid flow with boundary condition on the surface of the body with the shape corresponding to the given deformation.\\
These remarks may seem straightforward, but we encounter a crucial ambiguity trying to formulate the problem more specifically. Namely how can we specify the net motion of an object which is continuously changing shape? To quantify this motion it is necessary to attach a reference frame
to each unlocated shape. This is equivalent to choosing a standard location for each shape; more precisely to each unlocated shape there now corresponds a unique located shape. Once a choice of standard locations for shapes has been made, then we shall say that the rigid motion required to move between two different configurations is the displacement and rotation necessary to align their centers and axes. In what follows we shall develop a formalism, already used in   \cite{ShapereWilczek89} \cite{ShapereWilczek90}, which ensures us that the choice of axes for the unlocated shapes is completely arbitrary and that the rigid motion on the physical space is independent from this choice. 
This will be clear soon below.\\
For a given sequence of unlocated shapes $S_0(t)$, the corresponding sequence of located shapes $S(t)$ are related by 
\begin{equation}
\label{located_shapes}
S(t)=\mathcal{R}(t)S_0(t)
\end{equation}
where $\mathcal{R}$ is a rigid motion. This relation expresses how to recover the located shapes $S(t)$ given the unlocated ones, i.e. $S_0(t)$. 
It is clear that we are dealing with a fiber bundle: the located shapes $S(t)$ live on the big manifold $Q=SE(2)\times \mathcal{S}$ and the unlocated ones, $S_0(t)$, live on the base manifold obtained by the quotient of the manifold $Q$ by the plane euclidean group $SE(2)$, i.e $\mathcal{S}=Q/SE(2)$.

To make (\ref{located_shapes}) more explicit we introduce a matrix representation for the group of Euclidean motions, of which $\mathcal{R}$ is a member. A two dimensional rigid motion consisting of a rotation $R$ followed by a translation $d$ may be represented as a $3\times3$ matrix
\begin{equation}
\label{se2}
[R,d]=\begin{pmatrix}R&d\\0&1\end{pmatrix}
\end{equation}
where $R$ is an ordinary $2\times 2$ rotation matrix, $d$ is a 2 component column vector. This is the matrix representation of the plane euclidean group action $SE(2)$ on the manifold $Q$ where the located shapes $S(t)$ live on. \\
Now in considering the problem of self propulsion we shall assume that our swimmer has control over its form but cannot exert net forces and torques on itself. A swimming stroke is therefore specified by a time-dependent sequence of forms, or equivalently unlocated shapes $S_0(t)$. The located shape will then be expressed exactly by formula \eqref{located_shapes}.

Our problem of determining the net rigid motion of the swimmer thus resolves itself into the computation of $\mathcal{R}(t)$ given $S_0(t)$. In computing this displacement it is most convenient to begin with infinitesimal motions and to build up finite motions by integrating. So let us define the infinitesimal motion $A(t)$ by
\begin{equation}
\label{gauge_potential}
\frac{d\mathcal{R}}{dt}=\mathcal{R}\bigl(\mathcal{R}^{-1}\frac{d\mathcal{R}}{dt}\bigr)\equiv\mathcal{R}A
\end{equation}
In this formula we can recognize the differential equation corresponding to formula (\ref{group_diff}), from which we understand that $A$ take values in the Lie algebra of the plane euclidean group: $\mathfrak g=se(2)$.
For any given infinitesimal change of shape $A$, formula (\ref{gauge_potential}), describes the net overall translation and rotation which results. We can integrate it to obtain
\begin{equation}
\mathcal{R}(t_2)=\mathcal{R}(t_1)\bar{P}\exp\Bigl[\int_{t_1}^{t_2}A(t)\,dt\Bigr]
\end{equation}
where $\bar{P}$ denotes a reverse path ordering, known in literature as chronological series \cite{Agrachev12}:
$$
\bar{P}\exp\Bigl[\int_{t_1}^{t_2}A(t)\,dt\Bigr]=1+\int_{t_1<t<t_2}A(t)\,dt+\iint_{t_1<t^{'}<t<t_2} A(t)A(t^{'})\,dt\,dt^{'}+\cdots
$$
The assignment of center and axes can be arbitrary, so we should expect that physical results are independent of this assignment. How does this show up in our formalism? A change in the choice of centers and axes can equally well be thought of as a change (rigid motion) of the standard shapes, let us write
\begin{equation}
\label{change_reference}
\tilde{S}_0=\Omega(S_0)S_0
\end{equation}
The located shapes $S(t)$ being unchanged, (\ref{located_shapes}) requires us to define \cite{ShapereWilczek89} \cite{ShapereWilczek90}
$$
\tilde{\mathcal{R}}(t)=\mathcal{R}(t)\Omega^{-1}(S_0(t))
$$
From this, the transformation law of $A$ follow
\begin{equation}
\label{gauge_equations}
\tilde{A}=\Omega A \Omega^{-1}+\Omega\frac{d\Omega^{-1}}{dt}
\end{equation}
from which we can recognize the transformation laws (\ref{transformation_law}) of an Ehresmann connection called also \textbf{Gauge potential}. Our freedom in choosing the assignment of axes shows up as a freedom of gauge choice on the space of standard shapes. Accordingly the final relationship between physical shapes is manifestly independent of such choices.\\
Our aim will be to compute this gauge potential $A\in se(2)$ in function of the unlocated shapes $S_0$ that our swimmer is able to control.
\section{Swimming in an ideal fluid}
\label{Swimming_in_an_ideal_fluid}
We focus on a swimmer immersed in an ideal and incompressible fluid. The dynamical problem of its self propulsion has been reduced to the calculation of the gauge potential $A$. In our model we assume that the allowed motions, involving the same sequence of forms will include additional time-dependent rigid displacements. In other words the actual motion will be the composition of the given motion sequence $S_0(t)$ and rigid displacements.
\subsection{System of coordinates}
Let $(O,\vece_1,\vece_2)$ be a reference Galilean frame by which we identify the physical space to $\mathbb{R}^2$. At any time the swimmer occupies an open smooth connected domain $\mathcal{B}$ and we denote by $\mathcal{F}=\mathbb{R}^2\setminus\bar{\mathcal{B}}$ the open connected domain of the surrounding fluid. The coordinates in $(O,\vece_1,\vece_2)$ are denoted with $x=(x_1,x_2)^T$ and are usually called spatial coordinates. Let us call $(-x_2,x_1)^T=x^{\bot}$.\\
Attached to the swimmer, we define also a moving frame $(O^*,\vece_1^{*},\vece_2^{*})$. Its choice is made such that its origin coincides at any time with the center of mass of the body. This frame represents the choice of the axes in the space of unlocated shapes. As we have shown before, the computation of the net rigid motion of the swimmer due to shape changes is independent from this choice that accordingly is arbitrary. The fact that this frame has always its origin in the center of mass is a matter of convenience: indeed this choice, and others (see Remark \ref{minimal_kinetic_energy}), tell us that the body frame is the one in which the kinetic energy of the body is minimal \cite{LeviCivita20}.\\
We denote by $x^{*}=(x_1^{*},x_2^{*})^T$ the related so called body coordinates. In this frame and at any time the swimmer occupies a region $\mathcal{B}^{*}$ and the fluid the domain $\mathcal{F}^{*}:=\mathbb{R}^2\setminus\bar{\mathcal{B}}^{*}$.\\
We define also the \textit{computational space}, that is the Argand-Gauss plane which we will need only to perform explicit calculations, endowed with the frame $(\vecO,\vecE_1,\vecE_2)$ and in which the coordinates are denoted $z=(z_1,z_2)^T$. In this space $D$ is the unit disk and $\mathcal{O}:=\mathbb{R}^2\setminus\bar{D}$.

\begin{remark}[Minimal Kinetic Energy]
\label{minimal_kinetic_energy}
In order to simplify the calculations, since, as said in the previous section, the choice of the body frame is arbitrary, we use the one of the minimal kinetic energy. It is the one in which the velocity of the center of mass is null. This condition is clearly satisfied if  its origin coincides with the center of mass at any time. Moreover in this frame the angular momentum with respect to the body frame has to be null.
\end{remark}

\begin{remark}
\label{frame_choice}
The orientation of the body frame remains arbitrary and does not effect the fact that it is the frame of minimal kinetic energy. One of the most used conventions to define a possible orientation of such a system is to choose as axes the eigenvectors of the moment of inertia of our body. Obviously as we have said before this choice does not effect the located shape, since it is independent on the choice of the frame.
\end{remark}

\subsection{Shape  changes}
\textbf{Banach spaces of sequences}.
Inspired by \cite{MunnierChambrion10}, we denote any complex sequence by $\vecc:=(c_k)_{k\geq 1}$ where for any $k\geq 1$, $c_k:=a_k+ib_k\in\mathbb{C}$, $a_k,\,b_k\in\mathbb{R}$. Most of the complex sequences we will consider live in the Banach space
$$
\mathcal{S}:=\bigl\{(c_k)_{k\geq 1}\,:\,\sum_{k\geq 1}k(|a_k|+|b_k|)<+\infty\bigr\}
$$
endowed with its natural norm $\|\vecc\|_{\mathcal{S}}:=\sum_{k\geq 1}k(|a_k|+|b_k|)$. This space is continuously embedded in
$$
\mathcal{T}:=\bigl\{(c_k)_{k\geq 1}\,:\,\sup_{z\in\partial D}\bigl|\sum_{k\geq 1}kc_kz^k\bigr|<+\infty\bigr\}
$$
whose norm is $\|\vecc\|_{\mathcal{T}}:=\sup_{z\in\partial D}|\sum_{k\geq 1}kc_kz^k|$, where $D$ is the unit disk of the computational space. 

\begin{definition}
\label{D}x
We call $\mathcal{D}$ the intersection of the unit ball of $\mathcal{T}$ with the space $\mathcal{S}$.
\end{definition}
This space will play an important role in the description of the shape changes that will follow.\\Finally we introduce also the Hilbert space
$$
\mathcal{U}:=\bigl\{(c_k)_{k\geq 1}\,:\,\sum_{k\geq 1}k(|a_k|^2+|b_k|^2)<+\infty\bigr\}
$$
whose norm is $\|\vecc\|_{\mathcal{U}}:=\sqrt{\sum_{k\geq 1}k(|a_k|^2+|b_k|^2)}$.  According to Parseval's identity we have

$$
\sum_{k\geq 1}k|c_k|^2\leq\sum_{k\geq 1}k^2|c_k|^2=\frac{1}{2\pi}\int_0^{2\pi}\bigl(\sum_{k\geq 1}kc_k e^{-ik\theta}\bigr)^2\,d\theta\leq \sup_{z\in\partial D}\big|\sum_{k\geq 1}kc_kz^k\big|^2
$$
Therefore we have the following space inclusions
$$
\mathcal{S}\subset\mathcal{T}\subset\mathcal{U}
$$
We have introduced these spaces because they will be crucial in the description of the shape changes of the idealized swimmer.\\\\
\subsubsection{Description of the shape changes}
\begin{figure}[H]
	\centering
		\includegraphics[width=10cm]{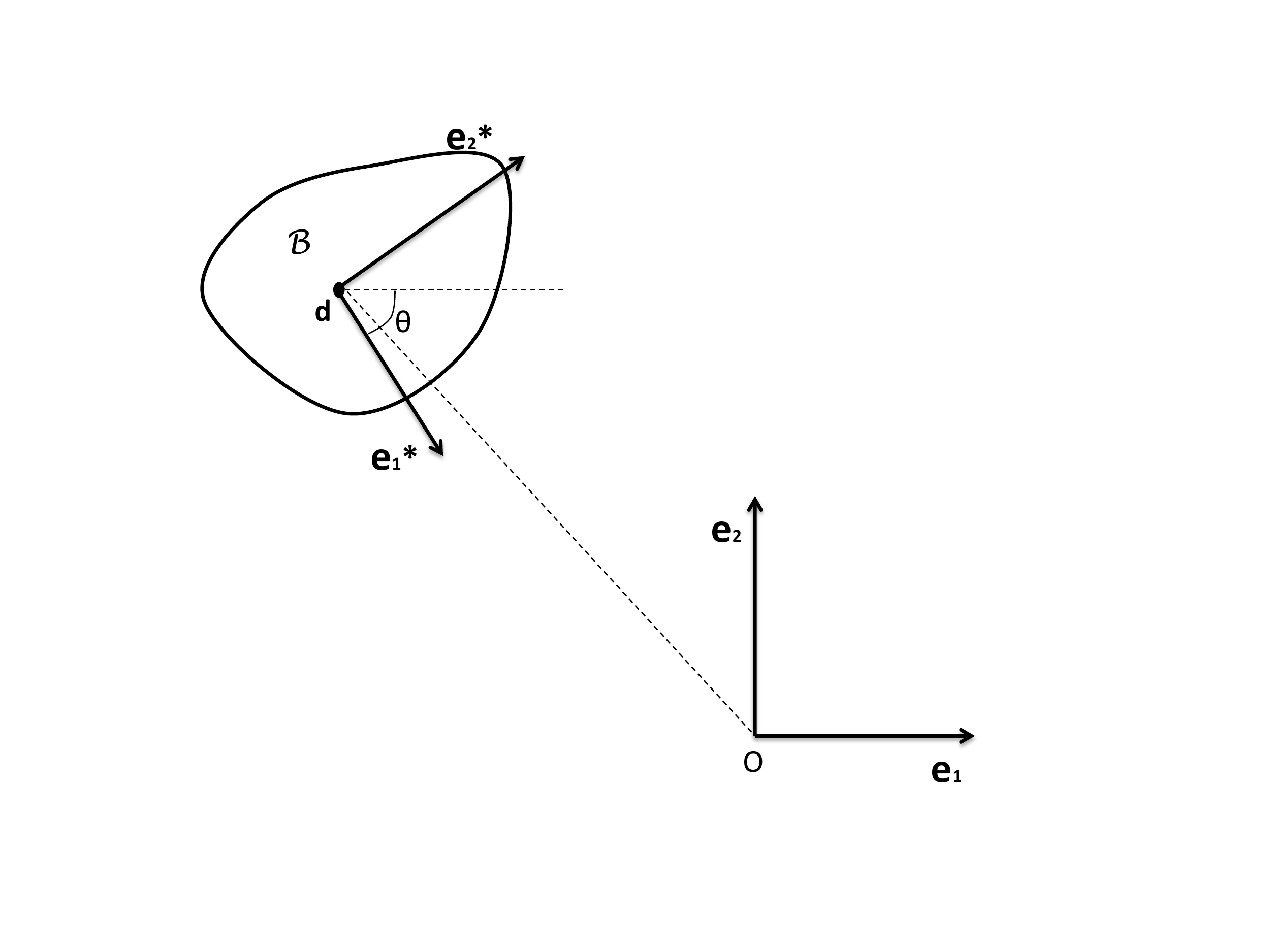}
	\caption{The physical space and the body frame.}
	\label{fig:frames}
\end{figure}

Following the line of thoughts of \cite{MunnierChambrion10} and \cite{MontgomeryStokes} the shape changes of the swimmer are described with respect to the moving frame $(O^{*},\vece_1^{*},\vece_2^{*})$ by a $\mathcal{C}^1$ diffeomorphism $\chi(\vecc)$, depending on a \textit{shape variable} $\vecc\in\mathcal{D}$ which maps the closed unit disk $\bar{D}$ of the computational space onto the domain $\mathcal{B}^{*}$ in the body frame. The diffeomorphisms $\chi(\vecc)$ allows us to associate to 
each sequence $\vecc$ a shape of the swimmer in the body frame. We can write, according to our notation, that for any $\vecc\in\mathcal{D}$ (see definition \ref{D}),

\begin{equation}
\chi(\vecc):\mathbb{C}\supset\bar{D}\rightarrow\mathbb{R}^2\equiv(O^{*},\vece_1^{*},\vece_2^{*})
\end{equation}
and $\bar{\mathcal{B}}^{*}=\chi(\vecc)(\bar{D})$.\\
We now explain how to build the map $\chi(\vecc)$ for any given sequence $\vecc$, see Fig \ref{fig:frames}.

  \begin{theorem}[Riemann Mapping Theorem]
  Let $\mathcal{K}$ be a simply connected open bounded subset of $\mathbb{C}$ with $0\in \mathcal{K}$. Then there exists an holomorphic isomorphism $f:D\rightarrow \mathcal{K}$ with $f(0)=0$.
  Any other isomorphisms with $f(0)=0$ are of the form $z\mapsto f(r z)$ with $r\in\partial D$ a rotation. All functions $f$ can be extended to an homeomorphism of $\bar{D}$ onto $\bar{\mathcal{K}}$ if and only if $\partial \mathcal{K}$ is a Jordan curve.
  \end{theorem}

  Defining $\mathbb{C}_{\infty}=\{\mathbb{C}\cup\infty\}$, if $\mathcal{O}=\mathbb{C}_{\infty}\setminus \bar{D}$, from the isomorphism $f$ we have also an isomorphism from $D$ to the exterior $\mathcal{F}^*$; we apply to $ \mathcal{F}^*$ the inversion $\rho(z):=\frac{1}{z}$ obtaining the open simply connected $G$, we find another Riemann- isomorphism $g:D\rightarrow G$ with $g(0)=0$. Then we consider $h=\rho\circ g=\frac{1}{g}:D\rightarrow \mathcal{F}^*$.
  The function $g$ is injective around zero, therefore $g^{'}(0)\neq 0$, it follows that $h$ has a pole of the first order in zero and therefore has  a Laurent expansion
  \begin{equation}
      h(z)=\frac{1}{z}+\frac{g^{''}(0)}{2}+\sum_{k=1}^{\infty}c_k z^k
  \end{equation}

  We now have the area theorem \cite{Pommerenke92}: if a function like $h$ is injective on the punctured disk then 
  we have 
  \begin{equation}\label{<1}
  \sum_{k=1}^{\infty}k|c_k|^2\leq 1
\end{equation}
If we want an isomorphism of $\mathcal{O}$ on $\mathcal{F}^{*}$ we take $\phi(\vecc)(z)=h(\frac{1}{z})$
\begin{equation}
\label{external_conformal_map}
\phi(\vecc)(z)=z+\sum_{k=1}^{\infty}\frac{c_k}{z^k}
\end{equation}
We now suppose that the boundary of $\mathcal{B}^*$ is a Jordan curve, i.e. simple closed curve in the plane, therefore the function $\phi(\vecc)$ can be extended to homeomorphism on the boundary. Now $\phi(\vecc):\bar{\mathcal{O}}\rightarrow \mathcal{F}^*$ can be extended continuously to all $\mathbb{C}_{\infty}$ setting in the interior of $D$

\begin{equation}
\label{deformation}
\chi(\vecc)(z):=z+\sum_{k\geq 1}c_k\bar{z}^k\,,\qquad(z\in\bar{D})
\end{equation}
Since $\bar{z}=\frac{1}{z}$ on $\partial D$ we deduce that the following map is continuous in $\mathbb{C}$ for all $\vecc\in\mathcal{D}$:
\begin{equation}
\Phi(\vecc)(z):=\begin{cases}
\chi(\vecc)(z)&\text{if\,\,} z\in D\\
\phi(\vecc)(z)&\text{if\,\,} z\in\bar{\mathcal{O}}=\mathbb{C}_{\infty}\setminus D
\end{cases}
\end{equation}
\begin{proposition}
\label{convergence}
For all $\vecc\in\mathcal{D}$, $\chi(\vecc):\bar{D}\rightarrow \bar{\mathcal{B}}^{*}$ and $\phi(\vecc):\bar{\mathcal{O}}\rightarrow \bar{\mathcal{F}}^{*}$ are both well defined and invertible. Further, $\chi(\vecc)|_{D}$ is a $\mathcal{C}^1$ diffeomorphism, $\phi(\vecc)|_{\mathcal{O}}$ is a conformal mapping and $\Phi(\vecc)$ is an homeomorphism form $\mathbb{C}$ onto $\mathbb{C}$.
\end{proposition}
\Proof
\cite{MunnierChambrion10}
\EndProof

\begin{remark}
\label{shape_circle}
Despite the generality of the Riemann Mapping Theorem, the way in which we decided to represent our diffeomorphism, lead us to some restrictions. Indeed in order to be sure that also $\chi(\vecc)$ is well defined -from proposition \ref{convergence}- we need to impose the restrictive condition $\vecc\in\mathcal{D}$, see \eqref{<1}, meaning that the shape variables have to be finitely bounded for both the norms of $\mathcal{S}$ and $\mathcal{T}$. To summarize we can say that to use the shape variable $\vecc\in\mathcal{D}$  allows us to describe all of the bounded non-empty connected shapes of the body that are not \textit{too far} from the unit ball.
\end{remark}

\subsection{Rigid motions}
The overall motion of our body in the fluid is, as said before, the composition of its shape changes with a rigid motion. The shape changes have been described in the previous subsection and, as we will see, the Gauge potential $A$ described at the beginning depends only on the shape variable $\vecc$  that is 
\begin{equation}
A=A(\vecc)
\end{equation} 
this will be evident in the next sections.\\
The net rigid motion is described by an element of the planar euclidean group as explained in subsection \ref{Gauge_potential}. More precisely it is given by a translation $\vecd$, which is the position of the center of mass, and a rotation $R$ of an angle $\theta$, that gives the orientation of the moving frame $(O^*,\vece_1^*,\vece_2^*)$ with respect to the physical one.\\
Let the shape changes be frozen for a while and consider a physical point $x$ attached to the body.  
Then there exists a smooth function $t\mapsto(\vecd(t),\theta(t))$ such that the point's coordinates in $(O,\vece_1,\vece_2)$ are given by $x=R(\theta)x_0+\vecd$. Next compute the time derivative expression $(\dot{\vecd},\dot{\theta})$. We deduce that the Eulerian velocity of the point is $\vecv_d(x)=\dot{\theta}(x-\vecd)^{\bot}+\dot{\vecd}$. It can be also expressed in the moving frame $(O^*,\vece_1^{*},\vece_2^{*})$ and reads $\vecv^{*}_d=\dot{\theta}(x^*)^{\bot}+\dot{\vecd}^{*}$ where
\begin{equation}
\label{reconstruction_eq}
\begin{pmatrix}\dot{\vecd}^*\\\dot\theta\end{pmatrix}=\mathcal{R}(\theta)^T\begin{pmatrix}\dot{\vecd}\\\dot\theta\end{pmatrix}
\end{equation}
where $\mathcal{R}$ is an element of the euclidean group $SE(2)$ of pure rotation.
\begin{remark}
Notice that  $\dot{\vecd}^*$ is not the time derivative of some $\vecd^*$ but only a symbol to expresse the velocity $\dot{\vecd}$ in the body frame.
\end{remark}

Let us return to the general case where the shape changes are taken into account. We deduce that the Eulerian velocity at a point $x$ of $\mathcal{B}$ is
$$
v(x)=\dot{\theta}(x-\vecd)^{\bot}+\dot{\vecd}+\mathcal{R}(\theta)\dot{\chi}(\vecc)[\chi(\vecc)^{-1}(R(\theta)^T(x-\vecd))]
$$
where the last term represent the velocity of deformation and is computed taking into account that $x=R^T(\theta)x^*=R^T(\theta)\chi(\vecc)(z)$.
When we express this velocity in the moving frame we get
\begin{equation}
\label{eulerian_velocity}
v^*(x^*)=(\dot{\theta}x^{*^{\bot}}+\dot{\vecd}^*)+\dot{\chi}(\vecc)(\chi(\vecc)^{-1}(x^*))\,,
\end{equation}
which is more compact and will be useful in what follows.
\subsection{Dynamics for ideal fluid}
 In this section we use some well known ideas developed for example in some works of A. Bressan \cite{Bressan07}, which further simplify the system of our idealized swimmer.\\
 We assume that the shape changes of our swimmer can be described by a finite number of shape parameters, i.e. $\vecc=(c_1,\cdots,c_m)$, thus we can call $\vecq=(q_1,\,\cdots,\,q_{m+3})=(\vecd,\theta,c_1,\cdots,c_m)$. This choice is widely spread in recent literature as in \cite{Bressan07,MasonBurdick99}, and implies that we focus only on a class of deformations which consist of particular shape changes that are sufficient to describe a wide 
 range of swimmer behaviors. Let us call $\tilde{\chi}$ the diffeomorphism which describes the superimposition of the shape changes with a rigid motion, more precisely $\tilde{\chi}(\vecq)(z):=[R(\theta),\vecd]\circ \chi(\vecc)(z)$  Assuming that there are no external forces, we wish to derive a system of equations describing the net motion of the body due to the shape changes and of the surrounding fluid expressed in the moving frame.
  Let $N=m+3$ and
\begin{equation}
\label{kinetic_body}
T(\vecq,\dot{\vecq})=\frac{1}{2}\sum_{i,j=1}^N A_{i\,j}(\vecq)\dot{q}^{i}\dot{q}^{j}
\end{equation}
describe the kinetic energy of the body. For simplicity, we assume that the surrounding fluid has unit density. Calling $v=v(x,t)$ its velocity at the point $x$, the kinetic energy of the surrounding fluid is given by
\begin{equation}
K=\int_{\mathcal{F}}\frac{|v(x)|^2}{2}\,dx
\end{equation}
  If the only active force is due to the scalar pressure $p$, the motion of the fluid is governed by the Euler equation for non-viscous, incompressible fluids:
\begin{equation}
\label{Euler}
v_{t}+v\cdot\nabla v=-\nabla p
\end{equation}
supplemented by the incompressibility condition $$\Div\,v=0.$$
In addition, we need a boundary condition
\begin{equation}
\label{boundary_condition}
\left\langle v-\sum_{k= 1}^N\frac{\partial\tilde{\chi}(\vecq)(z)}{\partial q_k}\dot{q}_k,\,-z\frac{\tilde{\chi}^{'}(\vecq)(z)}{|\tilde{\chi}^{'}(\vecq)(z)|}\right\rangle=0
\end{equation}
$-z\frac{\tilde{\chi}^{'}(\vecq)(z)}{|\tilde{\chi}^{'}(\vecq)(z)|}=n(x),\,(x=\tilde{\chi}(\vecq)(z),\,z\in\partial D)$  denotes the unit outer normal to the set $\tilde{\chi}(\vecc)(D)=\mathcal{B}$ at the point $x$, and is computed making the complex derivative of the function $\tilde{\chi}(\vecq)(z)$ expressed in polar coordinates that is
\begin{equation}
\label{normal}
n=i \frac{\partial_{\sigma}(\tilde{\chi}(\vecq)(e^{i\sigma}))}{|\partial_{\sigma}(\tilde{\chi}(\vecq)(e^{i\sigma))}|}=-e^{i\sigma}\frac{\tilde{\chi}^{'}(\vecq)(e^{i\sigma})}{|\tilde{\chi}^{'}(\vecq)(e^{i\sigma})|}=-z\frac{\tilde{\chi}^{'}(\vecq)(z)}{|\tilde{\chi}^{'}(\vecq)(z)|}
\end{equation}
which states that the velocity of the fluid has to be tangent to the surface of the body.
 To find the evolution of the coordinate $\vecq$, we observe that
\begin{equation}
\frac{d}{dt}\frac{\partial T}{\partial \dot{q_k}}=\frac{\partial T}{\partial q_k}+F_{k}\quad\quad k=1\cdots N
\end{equation}
where $T$ is the kinetic energy of the body and $F_i$ are the components of the external pressure forces acting on the boundary of $\mathcal{B}$.
 To determine these forces, we observe that, in connection with a small displacement of the $q^i$ coordinate, the work done by the pressure forces is
\begin{equation}
\delta W=-\delta q^{k}\cdot\int_{\partial D}\left\langle-z\frac{\tilde{\chi}^{'}(\vecq)(z)}{|\tilde{\chi}^{'}(\vecq)(z)|} ,\,\frac{\partial\tilde{ \chi}(\vecq)}{\partial q_k}(z)\right\rangle p(\tilde{\chi}(\vecq)(z))J(\vecq)(z)\,d\sigma
\end{equation}
The equation of motion are
\begin{equation}
\frac{d}{dt}\frac{\partial T}{\partial \dot{q^{k}}}=\frac{\partial T}{\partial q^{k}}-\int_{\partial D}\left\langle-z\frac{\tilde{\chi}^{'}(\vecq)(z)}{|\tilde{\chi}^{'}(\vecq)(z)|},\,\frac{\partial\tilde{ \chi}(\vecq)}{\partial q_{k}}(z)\right\rangle p(\tilde{\chi}(\vecq)(z))J(\vecq)(z)\,d\sigma
\end{equation}
We now show that, in the case of irrotational flow, the coupled system can be reduced to a finite dimensional impulsive Lagrangian system. It is well known (see \cite{Tucsnak10,MarchioroPulvirenti94}) that the velocity field of the fluid can be determined by setting $v=\nabla\psi$  and solving the Neumann problem in the exterior domain
\begin{equation}
\label{Laplace}
\begin{cases}
\Delta \psi=0 & \text{$x \in \mathcal{F}$}\\
n \cdot \nabla \psi=n\cdot v(x)|_{x=\tilde{\chi}(\vecq)(z)} &\text{$x \in \partial\mathcal{B}$}\\
|\psi|\rightarrow 0 & \text{$|x|\rightarrow\infty$}
\end{cases}
\end{equation}
where the boundary condition reads
\begin{equation}
\label{neumann_velocity}
n\cdot v(x)=-z\frac{\tilde{\chi}^{'}(\vecq)(z)}{|\tilde{\chi}^{'}(\vecq)(z)|}\cdot \sum_{k}\frac{\partial \tilde{\chi}(\vecq)}{\partial q_{k}}(z)\dot{q_{k}}
\end{equation}

Let us now consider the function $\tilde{\phi}(\vecq):\mathbb{R}^2\setminus D\rightarrow \mathbb{R}^2$ defined by the composition of $\phi(\vecq)(z)$  with the rigid motion $[R,\vecd]$, that clearly on the boundary of $D$ coincides with the function $\tilde{\chi}(\vecq)$.
From the linearity of (\ref{Laplace}) the solution will be linear in $\dot{\vecq}$.
\begin{equation}
\psi(z,\vecq,\dot{\vecq})=\sum_{k=1}^{N}\gamma_{k}(z,\vecq)\dot{q_{k}}
\end{equation}
The motion of the fluid can be obtained by solving the ordinary differential equation
\begin{equation}
\frac{d}{dt}\tilde{\phi}(\vecq)(z)=\frac{\partial \psi}{\partial x^L}(x,\vecq,\dot{\vecq})|_{x=\tilde{\phi}(\vecq)(z)}
\end{equation}
precisely
\begin{equation}
v_L(x,\vecq,\dot{\vecq})=\sum_{k}\frac{\partial \tilde{\phi}(\vecq)}{\partial q_{k}}(z)\dot{q}_{k}=\sum_{k}\frac{\partial\gamma_{k}}{\partial x^{L}}(x,\vecq)\dot{q_{k}}|_{x=\tilde{\phi}(\vecq)(z)}
\end{equation}
This has to be true for all curve $\mathbb{R}\ni t\longmapsto \vecq(t)$,  thus we have
\begin{equation}
\frac{\partial\tilde{ \phi}(\vecq)}{\partial q_{k}}(z)=\frac{\partial\gamma_{k}}{\partial x^{L}}(x,\vecq)|_{x=\tilde{\phi}(\vecq)(z)}
\end{equation}

We now prove that the term of the equations of motion relative to the pressure forces is a kinetic term

\begin{equation}
\begin{aligned}
\label{pressure_force}
F_{k}=&-\int_{\partial D}-z\frac{\tilde{\phi}^{'}(\vecq)(z)}{|\tilde{\phi}^{'}(\vecq)(z)|}\frac{\partial \tilde{\phi}(\vecq)}{\partial q_{k}}(z) p(\tilde{\phi}(\vecq)(z))J(\vecq)(z)\,d\sigma=\\
&-\int_{\partial D}-z\frac{\tilde{\phi}^{'}(\vecq)(z)}{|\tilde{\phi}^{'}(\vecq)(z)|}\frac{\partial\gamma_k}{\partial x^{L}}(\tilde{\phi}(\vecq)(z))p(\tilde{\phi}(\vecq)(z))J(\vecq)(z)\,d\sigma=\\ &
=-\int_{\partial \mathcal{B}}n^{L}(x)\frac{\partial\gamma_{k}}{\partial x^{L}}p\,dx
\end{aligned}
\end{equation}
applying the divergence theorem to (\ref{pressure_force})
\begin{align*}
&=\int_{x\in \mathcal{F}}\frac{\partial}{\partial x^{L}}(\frac{\partial\gamma_{k}}{\partial x^{L}}p)\,dx=\int_{x\in \mathcal{F}}\bigl(p\underbrace{\Delta\gamma_{k}}_{=0}+\nabla\gamma_{k}\cdot\nabla p\bigr)\,dx=\\&=\int_{x\in \mathcal{F}}\nabla\gamma_{k}\cdot\nabla p\,dx=\int_{z\in \mathbb{R}^{2}\setminus D}\frac{\partial \tilde{\phi}(\vecq)}{\partial q_{k}}\cdot\bigl(-v,_{t}-v\cdot\nabla v\bigr)\hat{J}(\vecq)(z)\,dz=\\&=-\int_{z\in \mathbb{R}^{2}\setminus D}\frac{\partial \tilde{\phi}(\vecq)}{\partial q_{k}}\cdot\frac{d}{dt}v\hat{J}(\vecq)(z)\,dz=\\ &
=-\int_{z\in \mathbb{R}^{2}\setminus D}[\frac{d}{dt}(\frac{\partial \tilde{\phi}(\vecq)_{L}}{\partial q_{k}}v_{L})-v_{L}\frac{\partial^{2}\tilde{\phi}(\vecq)}{\partial q_{j}\partial q^{k}}\dot{q}_{k}]\hat{J}(\vecq)(z)\,dz
\end{align*}
where $\hat{J}(\vecq)(z)$ is the determinant of the jacobian matrix of the function $\tilde{\phi}(\vecc)(z)$.
Let us define
\begin{equation*}
T^{f}=\frac{1}{2}\int_{x\in \mathcal{F}}|v|^{2}\,dx=\frac{1}{2}\sum_{i,j}^{N}\tilde{A}_{i\,j}\dot{q^{i}}\dot{q^{j}}
\end{equation*}
then
\begin{equation*}
F_{k}=-\Bigl(\frac{d}{dt}\int_{x\in \mathcal{F}}\frac{\partial|v|^{2}}{\partial \dot{q^{k}}}\,dx-\int_{x\in\mathcal{F}}\frac{\partial|v|^{2}}{\partial q^{k}}\,dx\Bigr)=-\frac{d}{dt}\frac{\partial T^{f}}{\partial \dot{q^{k}}}+\frac{\partial T^{f}}{\partial q^{k}}
\end{equation*}
In conclusion the system body+fluid is geodesic of Lagrangian
\begin{equation}
\label{kinetic_energy}
T=T^{body}+T^{f}
\end{equation}
In what follows for simplicity we will express all the quantities in the moving frame $(O^{*},\,\vece_1^*,\,\vece_2^*)$, denoting the total kinetic energy in this frame as $\stackrel{*}{T}$.
 We will now compute explicitly the Lagrangian. Let us start with the kinetic energy of the swimmer. Since we have chosen the body frame as the one of minimal kinetic energy, according to Konig theorem, there is a decoupling between the kinetic energy of the body due to its rigid motion and that due to its shape changes, recalling \eqref{kinetic_body}:
 \begin{equation}
 \label{Kinetic_body}
 \stackrel{*}{T}^{body}:=\frac{1}{2}m|\dot{\vecd}^*|^2+\frac{1}{2} I(\vecc)\dot{\theta}^2+\frac{1}{2}\int_{\mathcal{B}^*} \Bigl|\dot{\chi}(\vecc)(\chi(\vecc)^{-1}(x^*))\Bigr|^2\,dm^*
 \end{equation}
where $I(\vecc)$ is the moment of inertia of the body thought as rigid with frozen shape, and the last term being the kinetic energy of deformation. It can be computed as follows:
 $$
\int_{\mathcal{B}^*} \Bigl|\dot{\chi}(\vecc)(\chi(\vecc)^{-1}(x^*))\Bigr|^2\,dm^*=\int_D \bigl|\dot{\chi}(\vecc)(z)\bigr|^2\,dm_0=\pi\rho_0\sum_{k=1}^m\frac{|\dot{c}_k|^2}{k+1}\,
$$
where we used the formula \eqref{deformation} to compute the integral. Note that accordingly to remark (\ref{frame_choice}) the kinetic energy of the body in the frame $(O^*,\,\vece_1^*,\vece_2^*)$ does not depend on the orientation of the frame but only on its angular velocity.\\
 \subsubsection{Kinetic energy of the fluid}
Since we are interested on the effect of the shape changes of the swimmer on the fluid, in this subsection we will compute all the quantities in the body frame. As we have seen in subsection \ref{Gauge_potential} we can recover the rigid motion of the swimmer due to its deformation, exploiting the Gauge potential.\\
The kinetic energy of the fluid reads
 \begin{equation}
 \label{Kinetic_fluid}
\stackrel{*}{ T}^{f}:=\frac{1}{2}\int_{\mathcal{F}^*}|u^*|^2\,dm_f^*=\frac{1}{2}\int_{\mathcal{F}^*}|\nabla\psi^*|^2\,dm_f^*
 \end{equation}

 There $u^*=\nabla\psi^*$ and $\psi^*$ is the solution of the Neumann problem
 \begin{equation}
\label{Laplace2}
\begin{cases}
\Delta \psi^*=0 & \text{$x \in \mathcal{F}^*$}\\
n(x^*) \cdot \nabla \psi^*=n(x^*)\cdot v(x^*)|_{x^*=\chi(\vecq)(z)} &\text{$x^* \in \partial\mathcal{B}^*$}\\
|\psi^*|\rightarrow 0 & \text{$|x^*|\rightarrow\infty$}
\end{cases}
\end{equation}
 which is the same Neumann problem (\ref{Laplace}) expressed in the body frame. Indeed since the Laplacian operator is invariant under rototranslations, the function $\psi^*(x^*)=\psi([R(\theta),\vecd](x))$ is harmonic.

We will use complex analysis to compute the potential function $\psi^*$. We define the function $\xi(z):=\psi^*(\phi(\vecc)(z)),\,(z\in\mathcal{O})$, where $\psi^*$ is the potential function defined in (\ref{Laplace2}) expressed in the moving frame and recalling \eqref{external_conformal_map} $\phi(\vecc)(z)$ is the conformal map from $\bar{\mathcal{O}}=\mathbb{C}\setminus D$ to the external domain $\mathcal{F}^*$ . According to classical properties of conformal mappings, the function $\xi$ is harmonic in $\mathcal{O}$ and the following equality holds:
\begin{equation}
\frac{1}{2}\int_{\mathcal{F}^*}|\nabla\psi^*|^2\,dm_f^*=\frac{1}{2}\int_{\mathcal{O}}|\nabla\xi|^2\,dm_f^0
\end{equation}
The main advantage of this substitution is that $\xi$ is defined in the fixed domain $\mathcal{O}$, whereas $\psi^*$ was defined in $\mathcal{F}^*$ depending on $\vecc$.\\
In the moving frame is now easier to compute explicitly the boundary condition of the Neumann problem. The outer normal to $\partial\mathcal{B}^*$ is, recalling \eqref{normal}
\begin{equation}
n(x^*):=n_1(x^*)+in_2(x^*)=-z\frac{\phi^{'}(\vecc)(z)}{|\phi^{'}(\vecc)(z)|}\,
\end{equation}
where $\phi^{'}(\vecc)(z)$ is the complex derivative of $\phi(\vecc)$.
Recalling the following identity
\begin{equation}
\label{n_xi}
\frac{\partial_n\xi^r_j(z)}{|\phi^{'}(\vecc)(z)|}=\partial_n\psi^{*^{r}}_j(x^*)\qquad(x^*=\phi(\vecc)(z))
\end{equation}
and taking into account the expression (\ref{eulerian_velocity}) of $v^*$, we deduce that the Neumann boundary  condition \eqref{neumann_velocity} reads
\begin{equation}
\begin{aligned}
\partial_n\xi(z)=\nabla\xi\cdot n&=-\dot{d}_1^*\Re(z\phi^{'}(\vecc)(z))-\dot{d}_2^*\Im(z\phi^{'}(\vecc)(z))-\dot{\theta}\Im(\overline{\phi(\vecc)(z)}z\phi^{'}(\vecc)(z))\\
&-\Re(\overline{\dot{\chi}(\vecc)}z\phi^{'}(\vecc)(z))\,.
\end{aligned}
\end{equation}
This equality leads us to introduce the functions $\xi^r_j(\vecc)\,(j=1,2,3)$ and $\xi^d(\vecc)$ as being harmonic in $\mathcal{O}$ and satisfying the following Neumann boundary conditions:

\begin{align}
\label{normal_derivatives_state1}
&\partial_n\xi^r_1(\vecc)(z)=-\Re(z\phi^{'}(\vecc)(z)),\\
&\partial_n\xi^r_2(\vecc)(z)=-\Im(z\phi^{'}(\vecc)(z)),\\
\label{normal_derivatives_state3}
&\partial_n\xi^r_3(\vecc)(z)=-\Im(\overline{\phi(\vecc)(z)}z\phi^{'}(\vecc)(z)),\\
&\partial_n\xi^d(\vecc)(z)=-\Re(\overline{\dot{\chi}(\vecc)}z\phi^{'}(\vecc)(z)),\qquad (z\in\partial D).
\end{align}
In this way we spilt the harmonicity and the Neumann boundary conditions of the function $\xi$ into the same properties for the functions  $\xi^r_j(\vecc)\,(j=1,2,3)$ and $\xi^d(\vecc)$.\\
Next we have for all $\vecc\in\mathcal{D}$ 
$$
\Re(\overline{\dot{\chi}(\vecc)}z\phi^{'}(\vecc)(z))=\sum_{k=1}^m\dot{a}_k\Re(z^{k+1}\phi^{'}(\vecc)(z))+\dot{b}_k\Im(z^{k+1}\phi^{'}(\vecc)(z))
$$
%
\begin{proposition}[Potential decomposition]
According to the \textbf{Kirchhoff law} the 
following holds in the Sobolev space $H^1(\mathcal{O})$:
\begin{align}
\xi(\vecc)=\dot{d}_1^*\xi^r_1(\vecc)+\dot{d}_2^*\xi^r_2(\vecc)+\dot{\theta}\xi^r_3(\vecc)+\langle\xi^d(\vecc),\dot{\vecc}\rangle,
\end{align}
\end{proposition}
From the linearity of this expression with respect to $\dot{\vecd}^*,\,\dot{\theta},\,\dot{\vecc}$ and since the gradient function preserves the linearity, we deduce that the kinetic energy of the fluid is a quadratic function of $\dot{\vecd}^*,\,\dot{\theta},\,\dot{\vecc}$. 
\subsection{The Gauge potential and the equations of motion}
According to what proved in the preceding section the Lagrangian of our system is a quadratic form in $(\dot{\vecd}^*,\dot{\theta},\dot{\vecc})$, therefore it can be written in blocks as follows
\begin{equation}
\label{kinetic_energy1}
\stackrel{*}{T}(\dot{\vecd}^*,\dot{\theta},\dot{\vecc})=\frac{1}{2}\Bigl((\dot{\vecd}^{*^{T}},\dot{\theta})\mathbb{M}_r(\vecc)\begin{pmatrix}\dot{\vecd}^*\\\dot{\theta}\end{pmatrix}+2(\dot{\vecd}^{*^{T}},\dot{\theta})\mathbb{N}(\vecc)\dot{\vecc}+\dot{\vecc}^T\mathbb{M}_d(\vecc)\dot{\vecc}\Bigr)
\end{equation}
where $\mathbb{M}_r$, $\mathbb{N}$ and $\mathbb{M}_d$ play the role of the matrices $\mathcal{C}$  $\mathcal{M}$ and $\mathcal{B}$, introduced in the section \ref{impulse}, respectively
\begin{remark}
It is worth noting that in the physical space the kinetic energy is
$$T(\vecd,\theta,\vecc,\dot{\vecd},\dot{\theta},\dot{\vecc})$$
When it is expressed in the body frame instead it becomes
$$\stackrel{*}{T}(\vecc,\dot{\vecd}^*,\dot{\theta}, \dot{\vecc})$$
This expression does not depend on $\vecd$ and $\theta$ due to the symmetry of our model with respect to the position and orientation of the body in the fluid.\\
\end{remark}
As we have seen in the first section we are interested in determining the Gauge potential $A$ associated to our system which is
\begin{equation}
A=\mathcal{R}^{-1}\frac{d\mathcal{R}}{dt}=\begin{pmatrix}\dot{\theta}\begin{pmatrix}
0&1\\
-1&0
\end{pmatrix}&R(\theta)^T\begin{pmatrix}\dot{d}_1\\\dot{d}_2\end{pmatrix}\\
0\quad 0&0
\end{pmatrix}=\begin{pmatrix}\dot{\theta}\begin{pmatrix}
0&1\\
-1&0
\end{pmatrix}&\begin{pmatrix}\dot{d}_1^*\\\dot{d}_2^*\end{pmatrix}\\
0\quad 0&0
\end{pmatrix}
\end{equation}
Since all the matrices $\mathbb{M}_r$, $\mathbb{N}$ and $\mathbb{M}_d$ depend only on $\vecc$, the kinetic energy is independent from $\vecd$ and $\theta$ and the metric that it defines is \textbf{bundle like} (see Definition \ref{bundle_like}). In the principal fiber bundle $SE(2)\times\mathcal{S}\rightarrow \mathcal{S}$, the Gauge potential $A$  depends on the kinetic energy, through the equation of motion, therefore also $A$  does not depend on the state variables.\\
We now need to determine $(\dot{d}_1^*,\dot{d}_2^*,\dot{\theta})$. In order to do this
we compute the Hamiltonian associated to the Lagrangian function performing a partial legendre transformation on the $\dot{\vecq}^*$ variables.\\ Before passing to formal calculations we recall how to interpret the connection introduced before in the cotangent bundle setting following the steps presented in subsection \ref{impulse}. This construction was presented also in \cite{Cardin, marle91}. Let $VQ$ be the vertical subbundle and $V^{*}Q$ the dual of $VQ$. Denote with $p_{Q}:T^{*}Q\longrightarrow Q$ the cotangent projection and set $\tilde{\pi}:=\pi\circ p_Q$, $\tilde{\pi}:V^{*}Q\longrightarrow \mathcal{S}$. If $(\vecd,\theta,\vecc)$ are local fibered coordinates on $Q$, $(\vecd,\theta,\vecc,\vecp^*)$ are local fibered coordinates on $V^{*}Q$. Suppose that a control vector field  $\dot{\vecc}$ is given on $\mathcal{S}$ and that the path $\vecc(t)$ is an integral curve of $\dot{\vecc}$. Then the equation of motion are the local expression of a vector field $D_{\dot{\vecc}}$ over $V^{*}Q$ that projects on $\dot{\vecc}$ by $\tilde{\pi}$. Moreover the field $D_{\dot{\vecc}}$ is tangent to the fiber of $\tilde{\pi}$ only if the control is vanishing.\\
Recalling that $\vecq=(q_1,\,\cdots,\,q_{m+3})=(\vecd,\theta,\vecc)$
\begin{equation*}
\vecp^*=\Bigl(\frac{\partial  \stackrel{*}{T}}{\partial \dot{q}^*_i}\Bigr)_{i=1,2,3}=\mathbb{M}_r(\vecc)\begin{pmatrix}\dot{\vecd}^*\\\dot{\theta}\end{pmatrix}+\mathbb{N}(\vecc)\dot{\vecc}
\end{equation*}
which defines the translational and angular impulses of the system body plus fluid. From this we obtain
\begin{equation}
\label{Ham1}
\begin{pmatrix}\dot{\vecd^*}\\\dot{\theta}\end{pmatrix}=\mathbb{M}_r^{-1}(\vecc)\vecp^*-\mathbb{M}_r^{-1}(\vecc)\mathbb{N}(\vecc)\dot{\vecc}
\end{equation}


This expression is very convenient to study the motion of the shape-changing body since it gives the velocity with respect to the shape variable.

It is easy to recognize the terms of the sum in which the control equations are split according to Theorem \ref{splitting_control_eq} :
$$
X_{H_{0}}=\begin{pmatrix}\mathbb{M}_r^{-1}(\vecc)\vecp^*\\\underline{0}\end{pmatrix}\qquad\qquad X_{K_{\dot{\vecc}}}=0
$$
and
$$
hor(\dot{\vecc})=\begin{pmatrix}-\mathbb{M}_r^{-1}(\vecc)\mathbb{N}(\vecc)\dot{\vecc}\\\dot{\vecc}\end{pmatrix}
$$
Note  thus that we are exactly in the case of a \textbf{bundle-like metric} Therefore the equation of motion regarding the state variables are are exactly the ones given by formula \eqref{Ham1}.\\

To obtain the equation of motion regarding the conjugate variables, we follow the method explained in Lamb \cite{Lamb} and Munnier \cite{MunnierChambrion10}: we introduce
$\vecP$ and $\Pi$, the translational and angular impulses, as well as $\vecL$ and $\Lambda$, the
impulses relating to the deformations: 
\begin{equation}
\begin{aligned}
\begin{pmatrix}\vecP\\\Pi\end{pmatrix}=\mathbb{M}_r(\vecc)&\begin{pmatrix}\dot{\vecd}^*\\\dot{\theta}\end{pmatrix}&&\begin{pmatrix}\vecL\\\Lambda\end{pmatrix}=\mathbb{N}(\vecc)\dot{\vecc}\\
&\vecp^*=\begin{pmatrix}\vecP+\vecL\\\Pi+\Lambda\end{pmatrix}
\end{aligned}
\end{equation}

We start from the Lagrange equations
$$
\frac{d}{dt}\frac{\partial T}{\partial\dot{q}_i}-\frac{\partial T}{\partial q_i}=0,\qquad i=1,2,3
$$
and recall that $\stackrel{*}{T}(\vecc,\dot{\vecd}^*,\dot{\theta},\dot{\vecc})=T(\vecc,R(\theta)\dot{\vecd}^*,\dot{\theta},\dot{\vecc})$.\\
Therefore recalling that $\partial_{\theta}R(\theta)=R(\frac{\pi}{2})R(\theta)$
\begin{equation}
\begin{aligned}
&\frac{d}{dt}\frac{\partial T}{\partial\dot{\vecd}}-\frac{\partial T}{\partial \vecd}=\frac{d}{dt}\Bigl(\frac{\partial \stackrel{*}{T}}{\partial\dot{\vecd}^*}R(\theta)\Bigr)=\frac{d}{dt}(\vecP+\vecL)-\dot{\theta}(\vecP+\vecL)^{\bot}\\
&\frac{d}{dt}\frac{\partial T}{\partial\dot{\theta}}-\frac{\partial T}{\partial \theta}=\frac{d}{dt}(\frac{\partial \stackrel{*}{T}}{\partial\dot{\theta}})-R(\theta)^T\dot{\vecd}\cdot(\vecP+\vecL)^{\bot}=\\&\phantom{\frac{d}{dt}\frac{\partial T}{\partial\dot{\theta}}-\frac{\partial T}{\partial \vecd}}=\frac{d}{dt}(\Pi+\Lambda)-\dot{\vecd}^*\cdot(\vecP+\vecL)^{\bot}
\end{aligned}
\end{equation}
from these equations we get
\begin{equation}
\begin{aligned}
\frac{d}{dt}\begin{pmatrix}p_1^*\\p_2^*\end{pmatrix}+\dot{\theta}\begin{pmatrix}-p_2^*\\p_1^*\end{pmatrix}=0\\
\frac{d}{dt}p_3^*-\dot{d}_2^*p_1^*+d_1^*p_2^*=0
\end{aligned}
\end{equation}
Therefore the equation of motion in the body coordinates are
\begin{equation}
\label{Ham_reconstr}
\begin{cases}
\begin{pmatrix}\dot{\vecd^*}\\\dot{\theta}\end{pmatrix}=\mathbb{M}_r^{-1}(\vecc)\vecp^*-\mathbb{M}_r^{-1}(\vecc)\mathbb{N}(\vecc)\dot{\vecc}\\
\dot{p}_1^*=\dot{\theta}p_2^*\\
\dot{p}_2^*=-\dot{\theta}p_1^*\\
\dot{p}_3^*=\dot{d}_2^*p_1^*-\dot{d}_1^*p_2^*
\end{cases}
\end{equation}

Notice that these equations are exactly the ones presented in \cite{MasonBurdick99} which describe the evolution of the state and the conservation of the impulse.
\subsubsection{Simplification of the deformations}
Since we are interested in studying small deformation around a circular shape, see remark \ref{shape_circle}, in order to describe it we are interested only in expressing the distance from the origin of each point on circle's boundary in function of the shape parameters $\vecc=\veca+i\vecb$.
Let us consider deformation described by $m$ shape parameters.
According to formula (\ref{deformation}) the deformation can be written as:
$$
\chi(\vecc)(z)=z+\sum_{k=1}^m c_k \bar{z}^k\quad\text{for $z\in D$}
$$
therefore the modulus of a point on the boundary described in polar coordinates by $z=e^{i\sigma}$ is given by
\begin{equation}
\begin{aligned}
|\chi(\vecc)(z)|^2=&(z+\sum_{k=1}^m c_k \bar{z}^k)(\bar{z}+\sum_{k=1}^m \bar{c}_k z^k)=\\
&\Bigl(1+\sum_{k=1}^m(a_k\cos((k+1)\sigma)+b_k\sin((k+1)\sigma))\\
&+\sum_{h,k=1}^m(a_k+ib_k)(a_h-ib_h)\Bigr)
\end{aligned}
\end{equation}
taking the square root and using the Taylor expansion
around $\vecc=\underline{0}$ which corresponds to the circular shape we obtain
\begin{equation}
\label{radial_def}
|\chi(\vecc)(z)|=\bigl(1+\sum_{k=1}^m(\frac{a_k}{2}\cos((k+1)\sigma)+\frac{b_k}{2}\sin((k+1)\sigma))\bigr)+\sum_{k=1}^m o(c_k^2)
\end{equation}
where we can neglect all the terms of order grater or equal than 2 supposing $a_k,\,b_k$ small for all $k$, for example of order $\epsilon$.

For example in the case $m=2$ taking
\begin{align}
 &a_1=2\epsilon s_1&&b_1=0&&a_2=2\epsilon s_2&&b_2=2\epsilon s_3
\end{align}
we find exactly the formula for the swimmer deformation proposed by Mason and Burdick in (\cite{MasonBurdick99}).
\begin{equation}
\label{real_deformation}
F(\sigma,s)=\bigl[1+\epsilon(s_1\cos(2\sigma)+s_2\cos(3\sigma)+s_3\sin(3\sigma))\bigr]
\end{equation}
This will be the formula for the deformation that we will take from now on.
\begin{remark}
The fact that we are neglecting the terms of order $\epsilon^2$ is not in contrast with what said in \cite{MunnierChambrion10}. They consider that the points can rotate on the boundary at high frequency producing a macroscopic identical deformation, leading to different dynamics. Instead our type deformations \eqref{radial_def} do not allow this kind of behavior since it is a radial one. 
\end{remark}
We can express the equation of motion (\ref{Ham_reconstr}) using the real parameters $s_k$ as shape parameters instead of $c_k$ obtaining:
\begin{equation}
\begin{cases}
\label{real_motion_eq}
\begin{pmatrix}\dot{\vecd}^*\\\dot{\theta}\end{pmatrix}=\bar{\mathbb{M}}_r^{-1}(\vecs)\vecp^*-\bar{\mathbb{M}}_r^{-1}(\vecs)\bar{\mathbb{N}}(\vecs)\dot{\vecs}\\
\dot{p}_1^*=\dot{\theta}p_2^*\\
\dot{p}_2^*=-\dot{\theta}p_1^*\\
\dot{p}_3^*=\dot{d}_2^*p_1^*-\dot{d}_1^*p_2^*
\end{cases}
\end{equation}
where the matrices $\bar{\mathbb{M}}_r$ and $\bar{\mathbb{N}}$ have the same physical meaning of the matrices $\mathbb{M}_r$ and $\mathbb{N}$ but are expressed using the real shape parameters $\vecs$.\\
From now on we focus only on shape transformations near the identity, like \eqref{real_deformation} so that we can use real shape parameters to describe the deformation of the system. 

\subsubsection{Curvature of the connection: geometric and dynamic phase}
In this subsection we deal with the problem of having a net motion performing cyclical shape changes. Looking at equations \eqref{real_motion_eq}$_1$ is evident that there are two contributions: the one of $\overline{\mathbb{M}}_r^{-1}(\vecs)\vecp^*$ which involves the impulse and $-\overline{\mathbb{M}}_r^{-1}(\vecs)\overline{\mathbb{N}}(\vecs)\dot{\vecs}$ which is entirely geometrical.\\
\begin{itemize}
 \item $\boxed{\vecp^*=0}$\\\\
First, let us suppose that the system starts with zero initial impulse, i.e. $\vecp^*(0)=0$. With this assumption the last three equations of the system \eqref{real_motion_eq} have as unique solution the null one therefore, the first term of equation \eqref{real_motion_eq}$_1$ vanishes and the infinitesimal relationship between shape changes and body velocity is described by the local form of the connection computed above. Moreover we take into account the reconstruction relation \eqref{reconstruction_eq}, which links the state velocity expressed in the body frame with the one expressed in the physical frame,
\begin{equation}
\dot{g}=\begin{pmatrix}\dot{\vecd}\\\dot{\theta}\end{pmatrix}=-\begin{pmatrix}R(\theta)&0\\0&1\end{pmatrix}\overline{\mathbb{M}}_r^{-1}(\vecs)\overline{\mathbb{N}}(\vecs)\dot{\vecs}=-gA_i(\vecs)\dot{s}^i
\end{equation}
where $g$ is an element of the planar euclidean group $SE(2)$.
From these we recognize the expression of the Gauge potential \eqref{gauge_potential}.\\
We would like to find a solution for this equations that will aid in designing or evaluating motions that arise from shape variations. Because $SE(2)$ is a Lie group this solution will generally have the form
$$g(t)=g(0)e^{z(t)}$$
where $z\in se(2)$, the Lie algebra relative to $SE(2)$. An expansion for the Lie algebra valued function $z(t)$ is given by the Campbell-Hausdorff formula
\begin{equation}
z=\bar{A}+\frac{1}{2}\overline{[\overline{A},A]}+\frac{1}{3}\overline{[\overline{[\overline{A},A]},A]}+\frac{1}{12}[\overline{A},\overline{[\overline{A},A]}]+\cdots
\end{equation}
$$\overline{A}(t)\equiv \int_0^t A(\tau)\dot{\vecs}(\tau)\,d\tau$$
To obtain useful results in the spatial coordinates, examine the group displacement resulting from a periodic path $\alpha:[0,T]\rightarrow \mathbb{R}^m$, such that $\alpha(0)=\alpha(T)$. Taylor expand $A_i$ about $\alpha(0)$  and then regroup, simplify, apply integration by parts and use that the path is cyclic
\begin{equation}
z(\alpha)=-\frac{1}{2}F_{ij}(\alpha(0))\int_{\alpha}ds^i\,ds^j+\frac{1}{3}(F_{ij,k}-[A_i,F_{jk}])(\alpha(0))\int_{\alpha}ds^i\,ds^j\,ds^k+\cdots
\end{equation}
where
$$F_{ij}\equiv A_{j,i}-A_{i,j}-[A_i,A_j]$$
is called \textbf{curvature of the connection}.\\
For proportionally small deformations, the displacement experienced during one deformation cycle is:
\begin{equation}
g_{disp}=e^{z(\alpha)}\approx\exp\bigl(-\frac{1}{2}F_{ij}(\alpha(0))\int_{\alpha}ds^i\,ds^j\bigr)
\end{equation}
If the curvature $F$ is not null this displacement gives us the so called \textbf{geometric phase} that is the statement of the well-known Ambrose-Singer theorem \cite{AmbroseSinger}.\\
\item $\boxed{\vecp^*\neq 0}$\\\\
Let now suppose that the system starts with an initial impulse which is non zero. 
Thus the last three equations of \eqref{real_motion_eq} are not trivial. First of all we need to integrate this equations, which in function of the deformation $\vecs$ and $\dot{\vecs}$ take the form
\begin{equation}
\small
\label{eq_dotp}
\begin{cases}
\dot{p}^*_1=\bigl(\overline{\mathbb{M}}_r^{-1}(\vecs)\vecp^*\bigr)_3p_2^*-\bigl(A(\vecs)\dot{\vecs}\bigr)_3p_2^*\\
\dot{p}_2^*=-\bigl(\overline{\mathbb{M}}_r^{-1}(\vecs)\vecp^*\bigr)_3p_1^*+\bigl(A(\vecs)\dot{\vecs}\bigr)_3p_1^*\\
\dot{p}_3^*=\bigl(\overline{\mathbb{M}}_r^{-1}(\vecs)\vecp^*\bigr)_2p_1^*-\bigl(A(\vecs)\dot{\vecs}\bigr)_2p_1^*-\bigl(\overline{\mathbb{M}}_r^{-1}(\vecs)\vecp^*\bigr)_1p_2^*+\bigl(A(\vecs)\dot{\vecs}\bigr)_1p_2^*
\end{cases}
\end{equation}
these can be solved once the shape $\vecs$ is prescribed as a function of time, and as before we choose a periodic shape path  $\alpha:[0,T]\rightarrow \mathbb{R}^m$,with $\alpha(0)=\alpha(T)$.\\
Let us now consider the equation of motion regarding the state variables.  We have both the contributions: the geometrical one, already studied in the case with zero impulse, and also the one depending on the impulse $\vecp^*$.
\begin{equation}
\label{dynamic_phase}
 \dot{g}=g\Bigl(\overline{\mathbb{M}}_r^{-1}(s)\vecp^*-A_i(s)\dot{s}^i\Bigr)
\end{equation}

As before the integration of this term along $\alpha(t)$ gives
$$
g_{disp}=g(0)e^{z(t)}
$$
where 
\begin{equation}
 z=\overline{Z}+\frac{1}{2}\overline{[\overline{Z},Z]}+\frac{1}{3}\overline{[\overline{[\overline{Z},Z]},Z]}+\frac{1}{12}[\overline{Z},\overline{[\overline{Z},Z]}]+\cdots 
\end{equation}
$$
\overline{Z}:=\int_0^t\overline{\mathbb{M}}_r^{-1}(\tau)\vecp^*(\tau)-A(\tau)\dot{\vecs}(\tau)\,d\tau
$$
In order to see that $g_{disp}$ is effectively the sum of two contribution let us focus on the third equation of \eqref{dynamic_phase}. It is
$$
\dot{\theta}=\bigl(\overline{\mathbb{M}}_r^{-1}(\vecs)\vecp^*\bigr)_3-\bigl(A(\vecs)\dot{\vecs}\bigr)_3
$$
from this we can easly recognize two terms. The first one integrated along $\alpha$ is
\begin{equation}
\label{theta_dyn_phase}
\int_0^T \bigl(\overline{\mathbb{M}}_r^{-1}(\alpha(\tau))\vecp^*(\tau)\bigr)_3\,d\tau\,,
\end{equation} 
which value depends strictly on the evolution of the impulse $\vecp^*$ given by equations \eqref{real_motion_eq}$_{2-4}$. 
The second term is the geometric contribution analyzed in the previous section which depends on the curvature of the connection.\\
Once we have integrated this system and obtained the time evolution of $\theta$ we can solve also the ODEs regarding $\dot{\vecd}$ which are
\begin{equation}
 \dot{\vecd}=R(\theta)\bigl(\overline{\mathbb{M}}_r^{-1}(\vecs)\vecp^*\bigr)_{1,2}-R(\theta)\bigl(A(\vecs)\dot{\vecs}\bigr)_{1,2}
\end{equation}
Also for these two equations it is clear that there are two terms. One is always the geometric one, depending only on the shape $\vecs$ and $\dot{\vecs}$. The other one integrated over $\alpha$ gives
\begin{equation}
\label{state_dyn_phase}
\int_0^T R(\theta(\tau))\bigl(\overline{\mathbb{M}}_r^{-1}(\alpha(\tau))\vecp^*(\tau)\bigr)_{1,2}\,d\tau
\end{equation}
which is due to the presence of the impulse.\\
The two additional terms \eqref{theta_dyn_phase} and \eqref{state_dyn_phase}  are exactly the so called \textbf{dynamic phase} presented in section \ref{geometric_dynamic} and represent the gap on the fiber $(d_1,d_2,\theta)$ performed by the swimmer after a periodical change of shape.
\end{itemize}
\section{Controllability}
In this section we will focus on the controllability of our system, i.e. its ability to move everywhere in the plane changing its shape.\\
First of all we introduce some classical definition and results that will be useful in what follows
\subsection{Tools in geometric control theory}
Let us consider the following control system
\begin{equation}
\dot{q}=\mathcal{F}(q,u)
\label{control}
\end{equation}
where $q$ are local coordinates for smooth manifold $Q$ with $\dim Q=n$ and $u:[0,T]\rightarrow U\subset \mathbb{R}^{m}$ is the set of admissible controls.
The unique solution of (\ref{control}) at time $t\geq t_{0}$ with initial condition  $q(t_{0})=q_{0}$ and input function $u(\cdot)$ is denoted $q(t,t_{0},q_{0},u)$.
\begin{definition}
\begin{itemize}
\item
The \textbf{reachable set} $R^{V}(q_{0},T)$ is the set of points in $Q$ which are reachable from $q_{0}$ at exactly time $T>0$, following trajectories which, for $t\leq T$ remain in a neighborhood $V$ of $q_{0}$
\item
The system (\ref{control}) is \textbf{locally accessible} from $x_{0}$ if, for any neighborhood $V$ of $q_{0}$ and all $T>0$ the set $R^{V}_{T}(q_{0})=\bigcup_{t\leq T}R^{V}(q_{0},t)$ contains a non empty open set.
\item
The system (\ref{control}) is \textbf{locally strong accessible} from $q_{0}$ if for any neighborhood $V$ of $q_{0}$ and all $T>0$ sufficiently small, the set $R^{V}(q_{0},T)$ contains a non empty open set.
\item
The system (\ref{control}) is \textbf{controllable}, if for every $q_{1}\,,\,q_{2}\in Q$ exists a finite time $T>0$ and an admissible control $u:[0,T]\rightarrow U$ such that $q(T,0,q_{1},u)=q_{2}$
\end{itemize}
\end{definition}
Let now suppose the system \eqref{control} to be an affine non linear control system, namely
\begin{equation}
\dot{q}= \mathcal{F}(q,u)=f(q)+\sum^{m}_{j=1}g_{j}(q)u_{j}
\label{sist_non_lin_affine}
\end{equation}
We now present some general results for this type of control systems
\begin{definition}
The \textbf{strong accessibility algebra} $\mathcal{C}_{0}$ is the smallest subalgebra of the Lie algebra of smooth vector fields on $M$ containing the control vector fields $g_{1}\ldots g_{m}$, which is invariant under the drift vector field $f$, that is $[f,X]\in \mathcal{C}_{0},\,\forall X\in\mathcal{C}_{0}$, every element of the algebra $\mathcal{C}_{0}$ is a linear combination of repeated Lie brackets of the form $[X_{k},[X_{k-1},[\ldots,[X_{1},g_{j}]\ldots]]]$ for $j=1\ldots m$ and where $X_{i}\in\{f,g_{1},\ldots,g_{m}\}$.

The \textbf{strong accessibility distribution} $C_{0}$ is the corresponding involutive distribution $C_{0}(q)=\{X(q)|X\in\mathcal{C}_{0}\}$.
\label{sad}
\end{definition}

\begin{proposition}
Let $q_{e}$ be an equilibrium point of the system (\ref{sist_non_lin_affine}). The linearization of the system (\ref{sist_non_lin_affine}) at  $q_{e}$ is controllable if
\begin{equation}
rank\Bigl[g|\frac{\partial f}{\partial q}g|\bigl(\frac{\partial f}{\partial q}\bigr)^{2}g|\ldots|\bigl(\frac{\partial f}{\partial q}\bigr)^{n-1}g\Bigr]|_{q_{e}}=n
\end{equation}
\end{proposition}
We say that the Strong Accessibility Rank Condition at $q_{0}\in Q$ is satisfied if
\begin{equation}
\dim C_{0}(q_{0})=n
\end{equation}
\begin{proposition}
We say that the system (\ref{sist_non_lin_affine}) is locally strong accessible from $q_{0}$ if the strong accessibility rank condition is satisfied.
\end{proposition}

\begin{proposition}
\label{driftless_controllability}
If the system (\ref{sist_non_lin_affine}) is driftless, namely
\begin{equation}
\dot{q}=\sum_{i=1}^{m}u_{i}g_{i}(q)
\end{equation}
its controllability is equivalent to its strong accessibility.
\end{proposition}


Let us recall the definition of iterated Lie brackets \cite{Coron56}
\begin{definition}
Let $f\in C^{\infty}$ and $g\in C^{\infty}$ we define by induction on $k\in\mathbb{N}$ $ad^{k}_{f}g\in C^{\infty}$
\begin{align*}
&ad^{0}_{f}g:=g\\
&ad^{k+1}_{f}g:=[f,ad^{k}_{f}g],\,\forall k\in\mathbb{N}.
\end{align*}
\end{definition}
We are now ready to give a sufficient condition for small time local controllability

\begin{theorem}
\label{drift_not_zero}
Assume that the controlled vector fields $g_1\cdots g_m$ generate a Lie algebra $\mathcal{L}ie\{g_1\cdots g_m\}$ that satisfies $\mathcal{L}ie\{g_1\cdots g_m\}=T_q Q$ for all $q$ in $Q$ then the corresponding affine system
$$
\dot{q}=f(q)+\sum_{i=1}^m g_i(q)u_i
$$
is strongly controllable whenever there are no restrictions on the size of the controls.
\end{theorem}


\subsection{Swimmer controllability}
Let us consider the control system (\ref{Ham_reconstr}), since they involve the impulse $\vecp^*$  we have two different type of control system depending on the initial value of this impulse. If it is zero, we have a non linear drifltess affine control system, whose controllability can be proved with classical techniques, instead if it is not zero we have a non linear affine system with drift, which is more tricky to study.
\begin{remark}[Scallop Theorem]
 Note that in the case of zero initial impulse, if we have only one real shape parameter we are exactly in the case of the famous \textbf{Scallop Theorem} according to which if the swimmer performs a cyclical shape change $\alpha$ the net motion of the swimmer after a period is null.
 \begin{equation}
 \begin{pmatrix}\Delta \vecd\\\Delta\theta\end{pmatrix}=\int_0^T A(s(t))\dot{s}(t)\,dt=\int_{\alpha(0)}^{\alpha(T)}A(\alpha)\,d\alpha=0\quad\text{since $\alpha(0)=\alpha(T)$}
 \end{equation}
\end{remark}

Now let us study the controllability of this system in both cases of interest: $\vecp_0^*=0$ and $\vecp^*_0\neq 0$.
\subsubsection{Case $\boxed{\vecp^*(0)=0}$}
In this subsection we want to study the controllability of the system which starts with zero impulse. According to what said before this means that we deal with a non linear driftless affine control system.\\\\
\textbf{Case of $3$ real shape parameters}\\\\
In this section we study exactly the case of three real controls, then we will generalized the results obtained to a larger number of parameters.
More precisely, suppose that the deformation of our swimmer is governed by $s_1,s_2,s_3$ and according to \cite{MasonBurdick99} its shape is described in polar coordinates in the body frame by
\begin{equation}
\label{swimmer_shape3}
 F(\sigma,s)=1+\epsilon(s_1\cos(2\sigma)+s_2\cos(3\sigma)+s_3\sin(3\sigma))
\end{equation}
The perfect irrotational fluid has density $\rho$ and the potential $\psi^*$ can be determined solving the Laplace problem with Neumann boundary conditions following the steps described the preceding sections. \\
After that it is possible to compute the expression of the connection and the equation of motion

\begin{equation}
\small
\begin{split}
\label{control_eq}
\begin{pmatrix}\dot{\vecd}^*\\\dot{\theta}\\\dot{p}_1^*\\\dot{p}_2^*\\\dot{p}_3^*\\\dot{s_1}\\\dot{s}_2\\\dot{s}_3\end{pmatrix}=\begin{pmatrix}-(1-\mu)s_2\\-(1-\mu)s_3\\0\\0\\0\\-(1-\mu)(s_3p_1^*+s_2p_2^*)\\1\\0\\0\end{pmatrix}&\epsilon^2 u_1+\\&+\begin{pmatrix}-s_1\\0\\\frac{2\pi \rho s_3}{M}\\\frac{2\pi \rho s_3p_2^*}{M}\\-\frac{2\pi \rho s_3p_1^*}{M}\\s_1p_2^*\\0\\1\\0\end{pmatrix}\epsilon^2 u_2+\begin{pmatrix}0\\-s_1\\-\frac{2\pi \rho s_2}{M}\\-\frac{2\pi \rho s_2p_2^*}{M}\\\frac{2\pi \rho s_2p_1^*}{M}\\-s_1p_2^*\\0\\0\\1\end{pmatrix}\epsilon^2 u_3
\end{split}
\end{equation}
with $\mu=\frac{2\pi \rho}{M+\pi \rho}$ and $M$ the mass of our body.

Due to the change of variables (\ref{reconstruction_eq}), the equations of motion have to be supplemented with a so-called reconstruction equation allowing to recover $\vecd$ knowing $\theta$:

\begin{equation}
\small
\begin{split}
\label{control_eq2}
\begin{pmatrix}\dot{\vecd}\\\dot{\theta}\\\dot{p}_1^*\\\dot{p}_2^*\\\dot{p}_3^*\\\dot{s}_1\\\dot{s}_2\\\dot{s}_3\end{pmatrix}=&\begin{pmatrix}\mathcal{R}(\theta)\begin{pmatrix}-(1-\mu)s_2\\-(1-\mu)s_3\\0\end{pmatrix}\\0\\0\\-(1-\mu)(s_3p_1^*+s_2p_2^*)\\1\\0\\0\end{pmatrix}\epsilon^2 u_1+\\&+\begin{pmatrix}\mathcal{R}(\theta)\begin{pmatrix}-s_1\\0\\\frac{2\pi \rho s_3}{M}\end{pmatrix}\\\frac{2\pi \rho s_3p_2^*}{M}\\-\frac{2\pi \rho s_3p_1^*}{M}\\s_1p_2^*\\0\\1\\0\end{pmatrix}\epsilon^2 u_2+\begin{pmatrix}\mathcal{R}(\theta)\begin{pmatrix}0\\-s_1\\-\frac{2\pi \rho s_2}{M}\end{pmatrix}\\-\frac{2\pi \rho s_2p_2^*}{M}\\\frac{2\pi \rho s_2p_1^*}{M}\\-s_1p_2^*\\0\\0\\1\end{pmatrix}\epsilon^2 u_3
\end{split}
\end{equation}
\begin{theorem}
\label{control_3_p0}
The system (\ref{control_eq2}) is controllable.
\end{theorem}

\Proof
First of all note that system (\ref{control_eq2}) is clearly of the type
$$\dot{q}=\sum_{i=1}^3 g_i(\theta,\vecp^*,\vecs)u_i$$
Since  the initial impulses are zero it  is reduced to only six non trivial equations, indeed we easily have that
$$
\vecp^*(t)=0\quad\forall t
$$
is a solution of the equations regarding $\vecp^*$ \eqref{eq_dotp}.\\
Accordingly to theorem (\ref{driftless_controllability}) to prove the controllability
 it suffices to verify the Lie algebra rank condition, i.e $dim\bigl(\mathcal{L}ie\{g_i\}_{i=1,2,3}\bigr)=6$.
 We compute all the vector fields $g_i$ and the Lie brackets of the first order $[g_i,g_j]$ with $i\neq j$ (details in the Appendix) and compute their determinant
\begin{equation}
\small
\det\Bigl\{g_1,g_2,g_3,[g_1,g_2],[g_2,g_3],[g_1,g_3]\Bigr\}=\frac{4 \pi  \mu  \rho  \epsilon ^{18} \left(\mu  M-2 \pi  (\mu -1) \rho  \left(s_2^2+s_3^2\right)\right)}{M^2}
\end{equation}
which is not null except for values $s_2=s_3=0$. Since we can control the shape parameters we are always able to move from these configurations, and therefore cross the submaifolds defined by the equations $s_2=s_3=0$.\\
Thus we can conclude that $g_1,g_2,g_3,[g_1,g_2],[g_2,g_3],[g_1,g_3]$ are always linearly independent and $dim(Lie\{g_i,i=1,2,3\})=6$, which proves the controllability result.
\EndProof
\textbf{General Case: $m>3$}\\\\
In this subsection we deal with a generalization of the previous controllability result.
Suppose that the shape of the swimmer is described by $m$ real parameters $s_i,\, i=1\cdots m$, which define a transformation near to the identity, whose expression is a generalization of formula \eqref{swimmer_shape3}. Moreover recall that we are still in the assumption that the swimmer starts with zero initial impulse in body coordinates. In this case the equation of motion turn out to be
\begin{equation}
\label{control_eq_gen}
 \begin{pmatrix}
  \dot{x}\\\dot{y}\\\dot{\theta}\\\dot{s}_1\\\vdots\\\dot{s}_m
 \end{pmatrix}=\sum_{i=1}^m \tilde{g}_i u_i
\end{equation}
Note that also in this case, since the initial value of $\vecp^*$ is null, $\vecp^*(t)=0$ is still a solution and therefore $\vecp^*$ does not appear in the system.
We now investigate the controllability of the system \eqref{control_eq_gen}.

\begin{theorem}
\label{control_general}
 The system (\ref{control_eq_gen}) is controllable.
\end{theorem}
\Proof
First of all observe that if we keep constant and equal to zero the last $m-3$ controls, i.e. $u_i=0,\,i=4\cdots m$ the last $m-3$ equations gives us easily $s_i(t)\equiv 0\quad\forall t,\quad\forall m\geq 4$. This means that the shape of the swimmer  is actually described by only $3$ parameters.
Therefore the remaining control equations have to be the same of the ones obtained in the previous section with $m=3$. This implies that the first six components of the vectors $\tilde{g}_j|_{s_i\equiv 0 i=4\cdots m},\,j=1,2,3$ have to be equal to the vectors $g_i$ defined before. As a consequence we have that
\begin{equation}
 Lie\{\begin{pmatrix}g_i\\\underline{0}\end{pmatrix},\,i=1,2,3\} \subset Lie\{\tilde{g}_i,\,i=1\cdots m\}
\end{equation}
Moreover we have also that the vector space generated by the last $m-3$ vector fields $\tilde{g}_i$ evaluated at $s_i\equiv 0,\, i=1,2,3$ have to be contained in the Lie algebra generated by all the $\tilde{g}_i$, since they are some of the generators.
\begin{equation}
 span\{\tilde{g}_j|_{s_i\equiv 0,\, i=1,2,3},\,j\geq 4\}\subset Lie\{\tilde{g}_i,\,i=1\cdots m\}
\end{equation}
Furthermore we have also obviously that
\begin{equation}
  Lie\{\begin{pmatrix}g_i\\\underline{0}\end{pmatrix},\,i=1,2,3\} \cap span\{\tilde{g}_j|_{s_i\equiv 0,\, i=1,2,3},\,j\geq 4\}=\{\underline{0}\}
\end{equation}
This implies
\begin{equation}
\begin{split}
 dim\Bigl( Lie\{\tilde{g}_i,\,i=1\cdots m\}\Bigr)\geq&\underbrace{dim\Bigl( Lie\{\begin{pmatrix}g_i\\\underline{0}\end{pmatrix},\,i=1,2,3\}\Bigr)}_{=6}+\\&+\underbrace{dim\Bigl(span\{\tilde{g}_j |_{s_i\equiv 0,\, i=1,2,3},\,j\geq 4\}\Bigr)}_{\geq m-3}
 \end{split}
\end{equation}
where the first equality derives from the proof done before in the case $m=3$.\\
Thus finally we obtain that
\begin{equation}
  dim\Bigl( Lie\{\tilde{g}_i,\,i=1\cdots m\}\Bigr)\geq m+3
\end{equation}
which proves the controllability of the system.
\EndProof
\subsubsection{Case $\vecp^*_0\neq 0$}
Let us suppose that our deformable body has an initial constant impulse $\vecp^*_0$ that is not null. As a consequence our control system is a system with drift of dimension $m+6$.\\\\
\textbf{Case of $3$ shape parameters}\\\\
We start we the simplest case of three control shape parameters; 
Since we start with an initial impulse that is not null we have the following control system with drift
\begin{equation}
\small
\begin{split}
\label{control_eq_p}
&\begin{pmatrix}\dot{\vecd}^*\\\dot{\theta}\\\dot{p}_1^*\\\dot{p}_2^*\\\dot{p}_3^*\\\dot{s_1}\\\dot{s}_2\\\dot{s}_3\end{pmatrix}=\begin{pmatrix}\overline{\mathbb{M}}_r^{-1}(\vecs)\vecp^*\\(\overline{\mathbb{M}}_r^{-1}(\vecs)\vecp^*)_{3}p_2^*\\(\overline{\mathbb{M}}_r^{-1}(\vecs)\vecp^*)_{3}p_1^*\\(\overline{\mathbb{M}}_r^{-1}(\vecs)\vecp^*)_2 p_1^*-(\overline{\mathbb{M}}_r^{-1}(\vecs)\vecp^*)_1 p_2^*\\0\\0\\0\end{pmatrix}+\\&+\begin{pmatrix}-(1-\mu)s_2\\-(1-\mu)s_3\\0\\0\\0\\-(1-\mu)(s_3p_1^*+s_2p_2^*)\\1\\0\\0\end{pmatrix}\epsilon^2 u_1+\begin{pmatrix}-s_1\\0\\\frac{2\pi \rho s_3}{M}\\\frac{2\pi \rho s_3p_2^*}{M}\\-\frac{2\pi \rho s_3p_1^*}{M}\\s_1p_2^*\\0\\1\\0\end{pmatrix}\epsilon^2 u_2+\begin{pmatrix}0\\-s_1\\-\frac{2\pi \rho s_2}{M}\\-\frac{2\pi \rho s_2p_2^*}{M}\\\frac{2\pi \rho s_2p_1^*}{M}\\-s_1p_2^*\\0\\0\\1\end{pmatrix}\epsilon^2 u_3\,.
\end{split}
\end{equation}

Which taking into account the reconstruction equations becomes

\begin{equation}
\small
\begin{split}
\label{control_eq_p2}
\begin{pmatrix}\dot{\vecd}\\\dot{\theta}\\\dot{p}_1^*\\\dot{p}_2^*\\\dot{p}_3^*\\\dot{s}_1\\\dot{s}_2\\\dot{s}_3\end{pmatrix}=&\begin{pmatrix}\mathcal{R}(\theta)\overline{\mathbb{M}}_r^{-1}(\vecs)\vecp^*\\(\overline{\mathbb{M}}_r^{-1}(\vecs)\vecp^*)_{3}p_2^*\\(\overline{\mathbb{M}}_r^{-1}(\vecs)\vecp^*)_{3}p_1^*\\(\overline{\mathbb{M}}_r^{-1}(\vecs)\vecp^*)_2 p_1^*-(\overline{\mathbb{M}}_r^{-1}(\vecs)\vecp^*)_1 p_2^*\\0\\0\\0\end{pmatrix}+\\&+\begin{pmatrix}\mathcal{R}(\theta)\begin{pmatrix}-(1-\mu)s_2\\-(1-\mu)s_3\\0\end{pmatrix}\\0\\0\\-(1-\mu)(s_3p_1^*+s_2p_2^*)\\1\\0\\0\end{pmatrix}\epsilon^2 u_1+\begin{pmatrix}\mathcal{R}(\theta)\begin{pmatrix}-s_1\\0\\\frac{2\pi \rho s_3}{M}\end{pmatrix}\\\frac{2\pi \rho s_3p_2^*}{M}\\-\frac{2\pi \rho s_3p_1^*}{M}\\s_1p_2^*\\0\\1\\0\end{pmatrix}\epsilon^2 u_2\\&+\begin{pmatrix}\mathcal{R}(\theta)\begin{pmatrix}0\\-s_1\\-\frac{2\pi \rho s_2}{M}\end{pmatrix}\\-\frac{2\pi \rho s_2p_2^*}{M}\\\frac{2\pi \rho s_2p_1^*}{M}\\-s_1p_2^*\\0\\0\\1\end{pmatrix}\epsilon^2 u_3
\end{split}
\end{equation}


\begin{theorem}
\label{control_3_drift}
The system (\ref{control_eq_p2}) is strongly controllable if there are no restrictions on the size of the controls, except at least on submanifolds of co-dimension greater than one defined by $p_1^*=0$, $p_2^*=0$, $p_3^*= constant\neq 0$. Moreover these submanifolds are invariant and the control system restricted to them is strongly controllable  if there are no restrictions on the size of the controls.
\end{theorem}

\Proof
The system (\ref{control_eq_p2}) is clearly of the type
\begin{equation}
\label{drift_general}
\dot{q}=\vecf(q)+\sum_{i=1}^3\vecg_i(q)u_i
\end{equation}
Applying theorem (\ref{drift_not_zero}) to prove the strong controllability, we have to verify that the Lie algebra generated by the vector fields $\vecg_i$ has the same dimension of the tangent space, i.e $dim(Lie\{\vecg_i,i=1,2,3\})=9$.
Thus we compute the Lie brackets of zero, first and second order of the vectors $\vecg_i$ (the detailed expressions are in the Appendix).

The determinant of these vector fields is

\begin{equation}
\small
\begin{aligned}
&\det\Bigl\{\vecg_1,\vecg_2,\vecg_3,[\vecg_1,\vecg_2],[\vecg_1,\vecg_3],[\vecg_2,\vecg_3],[\vecg_1[\vecg_2,\vecg_3]],[\vecg_2[\vecg_2,\vecg_3]],[\vecg_3[\vecg_2,\vecg_3]]\Bigr\}=\\
&\frac{8192 }{M^{10}}\pi ^7 \mu  p_2^* \rho ^7 s_2^2 s_3^2 \epsilon ^{36} (M p_2^*-2 \pi  p_1^* \rho  s_2 s_3)\\
& \Bigl(
M^2 (p_1^* ((2 (\mu -3)
   \mu +3) s_2-\mu  s_3)+
   +p_2^* ((2 \mu -3) s_2+((9-4 \mu ) \mu -6) s_3))-\\
   &-2 \pi  (\mu -1) M \rho  \bigl(2 \mu  p_1^* s_2
   \left(s_2^2-2 s_3^2\right)-p_1^* (s_2+s_3) \left(4 s_2^2-3 s_2 s_3+s_3^2\right)-\\
   &-p_2^* s_3 \left(-2 \mu 
   s_2^2+s_2^2+s_3^2\right)\bigr)+8 \pi ^2 (\mu -1)^2 \rho ^2 s_2 s_3 \left(s_2^2+s_3^2\right) (p_2^* s_2-p_1^*
   s_3)\Bigr).
\end{aligned}
\end{equation}

This is not vanishing except the following cases

\begin{itemize}
\item $\boxed{s_2=0}$\\ This case is easily solved, indeed we are controlling $\dot{s}_2$, therefore we are always able to move from this configuration and cross the hypersurface $s_2=0$.
\item $\boxed{s_3=0}$\\  This case is solved exactly in the same way as the preceding one.
\item $\boxed{p_2^*=0}$\\ We want to prove that we always have a vector field that is non tangent to this hyper surface. For this purpose we compute the scalar product between the gradient of the determinant and each vector field $\vecg_i$ and see if it is non zero along the hyper surface $p_2^*=0$. Let us consider for example the scalar product with $\vecg_2$
\begin{equation*}
\small
\begin{aligned}
&(\left\langle \nabla \text{det}\{\vecg_1,\cdots\},\vecg_2\right\rangle |_{p_2^*=0}) M^{11}=\\&32768 \pi ^9 \mu  (p_1^*)^3 \rho ^9 s_2^3 s_3^4 \epsilon ^{38} \Bigl(M^2 \left(\left(-2 \mu ^2+6 \mu -3\right) s_2+\mu  s_3\right)+2 \pi  M   \rho  (2 \left(\mu ^2-3 \mu +2\right) s_2^3+\Bigr.\\&\Bigl.+s_2^2 (s_3-\mu  s_3)-2 \left(2 \mu ^2-3 \mu +1\right) s_2 s_3^2-(\mu -1)
  s_3^3)+8 \pi ^2 (\mu -1)^2 \rho ^2 s_2 s_3^2 \left(s_2^2+s_3^2\right)\Bigr)
\end{aligned}
\end{equation*}
this is null only in the following cases
\begin{itemize}
\item[a)] $s_2=0$\\ From which we are always able to move as we have seen before
\item[b)] $s_3=0$\\ From which we are always able to move as we have seen before
\item[c)] $p_1^*=0$\\ Looking at equations \eqref{control_eq_p} this case means that we are on the submanifold $p_1^*=0$, $p_2^*=0$, $p_3^*= const$ that has co-dimension 3. In this case the system cannot leave this submanifold, which is invariant. 
\item[d)] $s_2=f(s_3)$\\ Here $f$ is a suitable function of $s_3$ such that 
\begin{equation*}
\small
\begin{aligned}
&\Bigl(M^2 \left(\left(-2 \mu ^2+6 \mu -3\right) s_2+\mu  s_3\right)+2 \pi  M   \rho  (2 \left(\mu ^2-3 \mu +2\right) s_2^3+\\&+s_2^2 (s_3-\mu  s_3)-2 \left(2 \mu ^2-3 \mu +1\right) s_2 s_3^2-(\mu -1)
   s_3^3)+8 \pi ^2 (\mu -1)^2 \rho ^2 s_2 s_3^2 \left(s_2^2+s_3^2\right)\Bigr)=0
\end{aligned}
\end{equation*}
Since we are controlling both $s_2$ and $s_3$ we can always move from this configuration.
\end{itemize}
In any case the points in which this scalar product is zero define a submanifold of co-dimension grater than one that we are always able to bypass.
Therefore we have proved that the vector field $\vecg_2$ is non tangent to the hyper surface $p_2^*=0$, except in the case of $p_1^*=0$, thus we are able to move from it using suitable controls.
\item $\boxed{p_2^*=\frac{2\pi p_1^*\rho s_2 s_3}{M}}$\\ 
Like in the previous case we compute the scalar product of the gradient of the determinant with $\vecg_2$ and see that it is non null except on sub manifold of co-dimension grater than one, more precisely
\begin{equation*}
\small
\begin{aligned}
&(\left\langle \nabla \text{det},\vecg_2\right\rangle |_{p_2^*=\frac{2\pi p_1^*\rho s_2 s_3}{M}}) M^{14}=\\&65536 \pi ^9 \mu  p_1^{*^{3}} \rho ^9 s_2^3 s_3^4 \epsilon ^{38} \left(M^2+2 \pi ^2 \rho ^2 s_2^2 s_3^2\right) (M^3 \left(\left(2 \mu
   ^2-6 \mu +3\right) s_2-\mu  s_3\right)-\\&2 \pi  M^2 \rho  \left(2 \left(\mu ^2-3 \mu +2\right) s_2^3+(4-3 \mu )s_2^2 s_3+(4-3 \mu )
   s_2 s_3^2-(\mu -1) s_3^3\right)-\\&4 \pi ^2 (\mu -1) M \rho ^2 s_2 s_3^2 \left((4 \mu -3) _s2^2+(2 \mu -3) s_3^2\right)+16 \pi ^3
   (\mu -1)^2 \rho ^3 s_2^3s_3^2 \left(s_2^2+s_3^2\right))
\end{aligned}
\end{equation*}
this is null only in the following cases
\begin{itemize}
\item[$\circ$] $s_2=0$, $s_3=0$, $p_1^*=0$, $s_2=h(s_3)$ where $h$ is a suitable function of $s_3$ in which the scalar product vanishes.\\ These cases have been already faced and treated before.
\end{itemize}
\item $\boxed{p_2^*=\tilde{f}(p_1^*,s_2,s_3)}$\\
Where $\tilde{f}$ is a function that represents the value of $p_2^*$ in which the last factor of the determinant is null. Also in this case computing $\left\langle \nabla \text{det},\vecg_2\right\rangle |_{p_2^*=\tilde{f}(p_1^*,s_2,s_3)}$ we see that all the points in which it is null define sub manifolds of co-dimension grater than one defined by the equation $p_1^*=0$, $p_2^*=0$ and $p_3^*=const\neq0$.
\end{itemize}
Finally we can conclude that the vector fields $$\vecg_1,\vecg_2,\vecg_3,[\vecg_1,\vecg_2],[\vecg_1,\vecg_3],[\vecg_2,\vecg_3],[\vecg_1[\vecg_2,\vecg_3]],[\vecg_2[\vecg_2,\vecg_3]],[\vecg_3[\vecg_2,\vecg_3]]$$ are always linearly independent except the submanifolds defined by $p_{1,2}^*=0$, $p_3^*=const$, which proves the first part of the theorem.\\ These submanifolds are invariant and on them the dimension of the system reduces to $6$. If we restrict to them, we have still a system with drift, whose strong controllability can be proved verifying that the vectors $g_1,g_2,g_3,[g_1,g_2],[g_2,g_3],[g_1,g_3]$ are linearly independent (see Theorem \ref{drift_not_zero}). Therefore the system restricted to each of these invariant submanifolds is strongly controllable if there is no restrictions on the size of the controls.

\EndProof

\textbf{General case $m>3$}\\\\
In the case of initial impulse not zero, as we have said before we have a control affine system with drift of dimension $m+6$.
\begin{equation}
\label{control_eqp2_gen}
 \begin{pmatrix}
  \dot{d}_1^*\\\dot{d}_2^*\\\dot{\theta}\\\dot{p}_1^*\\\dot{p}_2^*\\\dot{p}_3^*\\\dot{s}_1\\\vdots\\\dot{s}_m
 \end{pmatrix}=\tilde{\vecf}+\sum_{i=1}^m\tilde{\vecg}_i u_i
\end{equation}

\begin{theorem}
The system (\ref{control_eqp2_gen}) is strongly controllable except at least on submanifolds of co-dimension greater than one defined by $p_1^*=0$, $p_2^*=0$, $p_3^*= constant\neq 0$. These submanifolds are invariant and the control system restricted to them is strongly controllable if there are no restrictions on the size of the controls.

\end{theorem}
\Proof
To prove the strong controllability of the system (\ref{control_eqp2_gen})  we exploit the theorem (\ref{drift_not_zero}). Thus the condition to prove is that the dimension of the Lie algebra genereted by the control vector fields $\tilde{\vecg}_i$ has dimension $m+6$.
The proof is similar to the proof of the previous theorem. Using the following facts:
\begin{itemize}
 \item $Lie\{\begin{pmatrix}\vecg_i\\\underline{0}\end{pmatrix},i=1,2,3\}\subset Lie\{\tilde{\vecg}_i,i=1,\cdots,m\}$
 \item $span\{\tilde{\vecg}_j|_{s_i\equiv 0,i=1,2,3},j\geq 4\}\subset Lie\{\tilde{\vecg}_i,i=1,\cdots,m\}$
 \item $Lie\{\begin{pmatrix}\vecg_i\\\underline{0}\end{pmatrix},i=1,2,3\}\cap span\{\tilde{\vecg}_j|_{s_i\equiv 0,i=1,2,3},j\geq 4\}=\{\underline{0}\}$
\end{itemize}
we deduce that
\begin{equation}
\begin{split}
 dim\Bigl( Lie\{\tilde{\vecg}_i,\,i=1\cdots m\}\Bigr)&\geq \underbrace{dim\Bigl( Lie\{\begin{pmatrix}\vecg_i\\\underline{0}\end{pmatrix},\,i=1,2,3\}\Bigr)}_{=9}+\\&+\underbrace{dim\Bigl(span\{\tilde{\vecg}_j |_{s_i\equiv 0,\, i=1,2,3},\,j\geq 4\}\Bigr)}_{\geq m-3}
 \end{split}
\end{equation}
Which proves that $ dim\Bigl( Lie\{\tilde{\vecg}_i,\,i=1\cdots m\}\Bigr)=m+6$, except at least on the same submanifold on which the condition is not satisfied even in the case $m=3$ that are of co-dimension greater than one defined by $p_1^*=0$, $p_2^*=0$, $p_3^*= constant\neq 0$. Again if we restrict ourselves on this invariant submanifolds, recalling that in Theorem \ref{control_general} we have proved that the dimension of the Lie algebra generated by the $\tilde{\vecg}_i$ is $m+3$, the condition of controllability of Theorem \ref{drift_not_zero} is satisfied. Therefore the system is strongly controllable on the invariant submanifolds if there are no restrictions on the size of the controls.
\\

\EndProof

\section*{Conclusions and perspectives}
In this paper we have investigated the geometric nature of the swimming problem of a $2$-dimensional deformable body immersed in an ideal irrotational fluid. \\
We faced a new problem: the study of the controllability properties of a dynamical system which can start with a non zero initial impulse. Reinterpreting the hydrodynamic forces exerted by the fluid on the body, as kinetic terms, and describing the shape changes with a finite number of parameters, we derive the equation of motion of the system. Using classical techniques in control theory we are able to gain some good results for the controllability of this kind of system.\\
If it starts with zero initial impulse we recover results present in the  literature. We are always able to find a suitable rate of deformation which makes the swimmer moving between two different fixed configurations. If instead the body starts with an initial impulse different from zero, the swimmer can self-propel in almost any direction if it can undergo shape changes without any bound on their velocity.\\
The fact that we take into account the presence of an initial impulse not null, and the analysis of the controllability of this system seems innovative and makes the study of the self-propulsion of deformable bodies in an ideal fluid more accurate and complete.\\
The approach described in this paper can be extended in a number of natural ways.
To begin with, we have restricted our attention to planar swimmers. The general 3-dimensional case is conceptually straightforward, even though the way of describing the shape changes should be different.\\
The study of bodies that change their shape using only a finite number of parameters is the initial point of a more complex study of controlling the deformation by diffeomorphisms.
Future work will also explore the optimal control problem associated to these kind of systems, especially in the case of non zero initial impulse. 
\newpage
\section*{Appendix}
The vector Fields $g_i$ and their Lie brackets of the first order mentioned in theorem \ref{control_3_p0} are
\begin{equation}
\small
\begin{aligned}
&g_1=\epsilon^2\begin{pmatrix}\mathcal{R}(\theta)\begin{pmatrix}-(1-\mu)s_2\\-(1-\mu)s_3\\0\end{pmatrix}\\1\\0\\0\end{pmatrix}&& g_2=\epsilon^2\begin{pmatrix}\mathcal{R}(\theta)\begin{pmatrix}-s_1\\0\\\frac{2\pi \rho s_3}{M}\end{pmatrix}\\0\\1\\0\end{pmatrix}\\ &g_3=\epsilon^2\begin{pmatrix}\mathcal{R}(\theta)\begin{pmatrix}0\\-s_1\\-\frac{2\pi \rho s_2}{M}\end{pmatrix}\\0\\0\\1\end{pmatrix}
\end{aligned}
\end{equation}

The Lie brackets generated by these vector fields are
\begin{equation}
 \begin{aligned}
  &[g_1,g_2]=\epsilon^4\left(
\begin{array}{c}
 \frac{2 \pi  s_2 s_3 (\mu -1) \rho  \sin (\theta )- \cos (\theta ) \left(M \mu -2 \pi  s_3^2 (\mu -1) \rho \right)}{M} \\
 \frac{ \sin (\theta ) \left(2 \pi  s_3^2 (\mu -1) \rho -M \mu \right)-2 \pi  s_2 s_3 (\mu -1) \rho  \cos (\theta )}{M} \\
 0 \\
 0 \\
 0 \\
 0
\end{array}
\right)\\&[g_1,g_3]=\epsilon^4\left(
\begin{array}{c}
 \frac{ \sin (\theta ) \left(M \mu -2 \pi   s_2^2 (\mu -1) \rho \right)-2 \pi  s_2 s_3 (\mu -1) \rho  \cos (\theta )}{M} \\
 -\frac{\cos (\theta ) \left(M \mu -2 \pi s_2^2 (\mu -1) \rho \right)+2 \pi s_2 s_3 (\mu -1) \rho  \sin (\theta )}{M} \\
 0 \\
 0 \\
 0 \\
 0
\end{array}
\right)\\&[g_2,g_3]=\epsilon^4\left(
\begin{array}{c}
 \frac{2 \pi s_1 \rho  (s_2 \sin (\theta )+s_3 \cos (\theta ))}{M} \\
 \frac{2 \pi  s_1 \rho  (s_3 \sin (\theta )-s_2 \cos (\theta ))}{M} \\
 -\frac{4 \pi \rho }{M} \\
 0 \\
 0 \\
 0
\end{array}
\right)
  \end{aligned}
\end{equation}

The vector fields that we need to compute the Lie algebra generated by $\vecg_i$ in theorem \ref{control_3_drift} are
\begin{equation}
\begin{aligned}
&\vecg_1=\epsilon^2\begin{pmatrix}\mathcal{R}(\theta)\begin{pmatrix}-(1-\mu)s_2\\-(1-\mu)s_3\\0\end{pmatrix}\\0\\0\\-(1-\mu)(s_3p_1^*+s_2p_2^*)\\1\\0\\0\end{pmatrix}&& \vecg_2=\epsilon^2\begin{pmatrix}\mathcal{R}(\theta)\begin{pmatrix}-s_1\\0\\\frac{2\pi \rho s_3}{M}\end{pmatrix}\\\frac{2\pi \rho s_3p_2^*}{M}\\-\frac{2\pi \rho s_3p_1^*}{M}\\s_1p_2^*\\0\\1\\0\end{pmatrix}\\& \vecg_3=\epsilon^2\begin{pmatrix}\mathcal{R}(\theta)\begin{pmatrix}0\\-s_1\\-\frac{2\pi \rho s_2}{M}\end{pmatrix}\\-\frac{2\pi \rho s_2p_2^*}{M}\\\frac{2\pi \rho s_2p_1^*}{M}\\-s_1p_2^*\\0\\0\\1\end{pmatrix}
\end{aligned}
\end{equation}
Their Lie brackets of the first order are
\begin{equation}
\small
 \begin{aligned}
  &[\vecg_1,\vecg_2]=\epsilon^4\left(
\begin{array}{c}
 \frac{2 \pi   s_2 s_3 (\mu -1) \rho  \sin (\theta )- \cos (\theta ) \left(M \mu -2 \pi   s_3^2 (\mu -1) \rho \right)}{M} \\
 \frac{ \sin (\theta ) \left(2 \pi   s_3^2 (\mu -1) \rho -M \mu \right)-2 \pi  s_2 s_3 (\mu -1) \rho  \cos (\theta )}{M} \\
 0 \\
 0 \\
 0 \\
 \frac{2 \pi   s_3 (\mu -1) \rho  (p_1^* s_2-p_2^* s_3)-M p_2^*  (\mu -2)}{M} \\
 0 \\
 0 \\
 0
\end{array}
\right)\\&[\vecg_1,\vecg_3]=\epsilon^4\left(
\begin{array}{c}
 \frac{ \sin (\theta ) \left(M \mu -2 \pi  s_2^2 (\mu -1) \rho \right)-2 \pi   s_2 s_3 (\mu -1) \rho  \cos (\theta )}{M} \\
 -\frac{ \cos (\theta ) \left(M \mu -2 \pi   s_2^2 (\mu -1) \rho \right)+2 \pi   s_2 s_3 (\mu -1) \rho  \sin (\theta )}{M} \\
 0 \\
 0 \\
 0 \\
 \frac{ \left(2 \pi   s_2 (\mu -1) \rho  (p_1^* s_2+p_2^* s_3)-M (p_1^* (\mu -1)+p_2^*)\right)}{M} \\
 0 \\
 0 \\
 0
\end{array}
\right)\\&[\vecg_2,\vecg_3]=\epsilon^4\left(
\begin{array}{c}
 \frac{2 \pi   s_1 \rho  (s_2 \sin (\theta )+s_3 \cos (\theta ))}{M} \\
 \frac{2 \pi   s_1 \rho  (s_3 \sin (\theta )-s_2 \cos (\theta ))}{M} \\
 -\frac{4 \pi   \rho }{M} \\
 -\frac{4 \pi   \rho  \left(M p_2^*-2 \pi  p_1^*  s_2 s_3 \rho \right)}{M^2} \\
 -\frac{8 \pi ^2 p_2^*  s_2 s_3 \rho ^2}{M^2} \\
 \frac{2 \pi  p_1^*  s_1 \rho  (s_2+s_3)}{M} \\
 0 \\
 0 \\
 0
\end{array}
\right)
 \end{aligned}
\end{equation}

Finally the only non zero brackets of the second order are
\begin{equation}
\small
\begin{aligned}
&[\vecg_1,[\vecg_2,\vecg_3]]=\epsilon^6\left(
\begin{array}{c}
 -\frac{2 \pi  (2 \mu -3) \rho  (s_2 \sin (\theta )+s_3 \cos (\theta ))}{M} \\
 \frac{2 \pi  (2 \mu -3) \rho  (s_2 \cos (\theta )-s_3 \sin (\theta ))}{M} \\
 0 \\
 0 \\
 0 \\
 \frac{2 \pi   \rho  \left(M (p_1^* (s_2+s_3)+2 p_2^* s_3 (\mu -1))+4 \pi  s_2 s_3 (\mu -1) \rho  (p_2^*
   s_2-p_1^* s_3)\right)}{M^2} \\
 0 \\
 0 \\
 0
\end{array}
\right)\\
&[\vecg_2,[\vecg_2,\vecg_3]]=\epsilon^6\left(
\begin{array}{c}
 \frac{2 \pi   s_1 \rho  \left(\sin (\theta ) \left(3 M-2 \pi   s_3^2 \rho \right)+2 \pi   s_2 s_3 \rho  \cos (\theta
   )\right)}{M^2} \\
 \frac{2 \pi   s_1 \rho  \left(2 \pi   s_3 \rho  (s_2 \sin (\theta )+s_3 \cos (\theta ))-3 M \cos (\theta )\right)}{M^2} \\
 0 \\
 \frac{16 \pi ^2  s_3 \rho ^2 \left(M p_1^*+2 \pi  p_2^*  s_2 s_3 \rho \right)}{M^3} \\
 -\frac{16 \pi ^2  s_3 \rho ^2 \left(M p_2^*-2 \pi  p_1^*  s_2 s_3 \rho \right)}{M^3} \\
 \frac{2 \pi   s_1 \rho  \left(M p_1^*+2 \pi  p_2^*  s_3 \rho  (3 s_2+s_3)\right)}{M^2} \\
 0 \\
 0 \\
 0
\end{array}
\right)\\
&[\vecg_3,[\vecg_2,\vecg_3]]=\epsilon^6\left(
\begin{array}{c}
 \frac{2 \pi   s_1 \rho  \left(\cos (\theta ) \left(3 M-2 \pi   s_2^2 \rho \right)+2 \pi   s_2 s_3 \rho  \sin (\theta
   )\right)}{M^2} \\
 \frac{2 \pi   s_1 \rho  \left(3 M \sin (\theta )-2 \pi   s_2 \rho  (s_2 \sin (\theta )+s_3 \cos (\theta ))\right)}{M^2} \\
 0 \\
 \frac{16 \pi ^2  s_2 \rho ^2 \left(M p_1^*-2 \pi  p_2^*  s_2 s_3 \rho \right)}{M^3} \\
 -\frac{16 \pi ^2  s_2 \rho ^2 \left(M p_2^*-2 \pi  p_1^*  s_2 s_3 \rho \right)}{M^3} \\
 \frac{2 \pi   s_1 \rho  \left(M p_1^*-2 \pi  p_2^*  s_2 \rho  (s_2+3 s_3)\right)}{M^2} \\
 0 \\
 0 \\
 0
\end{array}
\right)
\end{aligned}
\end{equation}

\end{document}